\documentclass[journal]{IEEEtran}
\usepackage{hyperref}
\usepackage{ragged2e}
\usepackage{graphicx}
\usepackage{amsmath}
\usepackage{amssymb}
\usepackage{algorithm}
\usepackage{algorithmic}
\usepackage{array}
\usepackage{booktabs}
\usepackage{url}

\usepackage{cite}
\usepackage{multirow}
\usepackage{tabularx}
\usepackage{makecell}
\newcommand{\deferfloatstonextpage}{%
  \setcounter{topnumber}{0}%
  \suppressfloats[t]%
  \AddToHookNext{shipout/after}{\setcounter{topnumber}{4}}%
}
\usepackage{xcolor}
\usepackage{tikz}
\usetikzlibrary{patterns,patterns.meta}
\newcommand{\cmark}{\ensuremath{\checkmark}}
\newcommand{\pmark}{\ensuremath{\triangle}}
\newcommand{\dmark}{\ensuremath{--}}

\graphicspath{{DeLoop/}{./}}
\DeclareGraphicsExtensions{.pdf,.png,.jpg,.jpeg}
\definecolor{deloopblue}{RGB}{111,127,143}
\definecolor{langgraphorange}{RGB}{238,152,91}
\definecolor{fcreflectionteal}{RGB}{139,188,188}
\definecolor{autogenmustard}{RGB}{183,161,74}
\definecolor{ecdfnavy}{RGB}{43,59,146}
\definecolor{ecdfteal}{RGB}{126,193,190}
\definecolor{ecdfmint}{RGB}{189,223,187}
\definecolor{ecdfsteel}{RGB}{84,159,191}
\definecolor{breakdowncontrol}{RGB}{169,205,186}
\definecolor{breakdowntool}{RGB}{105,167,178}
\definecolor{breakdownwait}{RGB}{72,122,162}
\definecolor{llmcallgreen}{RGB}{191,223,210}
\definecolor{toolcallblue}{RGB}{37,125,139}
\definecolor{iterdepthorange}{RGB}{186,86,50}
\definecolor{successpurple}{RGB}{92,51,139}
\newcommand{\legendbox}[1]{\raisebox{0.12ex}{\textcolor{#1}{\rule{1.05em}{0.55em}}}}
\newcommand{\patternbox}[2]{%
  \raisebox{0.12ex}{\begin{tikzpicture}[baseline=0pt,line width=0.25pt]
    \path[fill=#1] (0,0) rectangle (1.05em,0.55em);
    \path[pattern={#2}] (0,0) rectangle (1.05em,0.55em);
    \draw[black] (0,0) rectangle (1.05em,0.55em);
  \end{tikzpicture}}}
\newcommand{\hatchbox}[1]{\patternbox{#1}{Lines[angle=45,distance=1.5pt,line width=0.25pt]}}
\newcommand{\backhatchbox}[1]{\patternbox{#1}{Lines[angle=-45,distance=1.5pt,line width=0.25pt]}}
\newcommand{\dotbox}[1]{\patternbox{#1}{Dots[distance=1.6pt,radius=0.22pt]}}
\newcommand{\crossbox}[1]{%
  \raisebox{0.12ex}{\begin{tikzpicture}[baseline=0pt,line width=0.25pt]
    \path[fill=#1] (0,0) rectangle (1.05em,0.55em);
    \path[pattern={Lines[angle=45,distance=1.5pt,line width=0.25pt]}] (0,0) rectangle (1.05em,0.55em);
    \path[pattern={Lines[angle=-45,distance=1.5pt,line width=0.25pt]}] (0,0) rectangle (1.05em,0.55em);
    \draw[black] (0,0) rectangle (1.05em,0.55em);
  \end{tikzpicture}}}
\newcommand{\legendline}[2][0.42pt]{\raisebox{0.42ex}{\textcolor{#2}{\rule{1.15em}{#1}}}}
\newcommand{\markerline}[3][0.9pt]{%
  \legendline[#1]{#2}\hspace{-1.15em}\makebox[1.15em][c]{\textcolor{#2}{#3}}}
\newcommand{\panelcaption}[1]{\par\vspace{0.3mm}{\fontsize{6.2}{7.2}\selectfont #1}}
\newsavebox{\ecdfimagebox}
\newsavebox{\breakdownimagebox}
\newcommand{\tokenmethodlegend}{%
  {\fontsize{6.2}{7.2}\selectfont\fontfamily{phv}\selectfont
  \hatchbox{langgraphorange}\hspace{0.35em}LangGraph\hspace{1.25em}
  \dotbox{fcreflectionteal}\hspace{0.35em}FC-Reflection\hspace{1.25em}
  \backhatchbox{autogenmustard}\hspace{0.35em}AutoGen\hspace{1.25em}
  \crossbox{deloopblue}\hspace{0.35em}AgentLoop}}
\newcommand{\ftrmethodlegend}{%
  {\fontsize{5.8}{7}\selectfont\fontfamily{phv}\selectfont
  \hatchbox{langgraphorange}\hspace{0.35em}LangGraph\hspace{1.2em}
  \dotbox{fcreflectionteal}\hspace{0.35em}FC-Reflection\hspace{1.2em}
  \backhatchbox{autogenmustard}\hspace{0.35em}AutoGen\hspace{1.2em}
  \crossbox{deloopblue}\hspace{0.35em}AgentLoop}}
\newcommand{\ecdfmethodlegend}{%
  {\fontsize{6.2}{7.2}\selectfont\fontfamily{phv}\selectfont
  \markerline{ecdfnavy}{\scalebox{1.15}{$\blacklozenge$}}\hspace{0.35em}AgentLoop\hspace{1.25em}
  \markerline{ecdfteal}{\scalebox{1.9}{$\bullet$}}\hspace{0.35em}LangGraph\hspace{1.25em}
  \markerline{ecdfmint}{\scalebox{1.4}{$\blacktriangle$}}\hspace{0.35em}FC-Reflection\hspace{1.25em}
  \markerline{ecdfsteel}{\scalebox{1.15}{$\bigstar$}}\hspace{0.35em}AutoGen}}
\newcommand{\breakdownlegend}{%
  {\fontsize{6.2}{7.2}\selectfont\fontfamily{phv}\selectfont
  \hatchbox{breakdowncontrol}\hspace{0.35em}Model/Control\hspace{1.25em}
  \dotbox{breakdowntool}\hspace{0.35em}Tool Execution\hspace{1.25em}
  \crossbox{breakdownwait}\hspace{0.35em}Framework waiting}}
\newcommand{\resourcedepthlegend}{%
  {\fontsize{6.2}{7.2}\selectfont\fontfamily{phv}\selectfont
  \legendbox{llmcallgreen}\hspace{0.35em}LLM Calls\hspace{1.15em}
  \legendbox{toolcallblue}\hspace{0.35em}Tool Calls\hspace{1.15em}
  \legendline{iterdepthorange}\hspace{-0.48em}\textcolor{iterdepthorange}{$\blacklozenge$}\hspace{0.30em}Iter Depth\hspace{1.15em}
  \legendline{successpurple}\hspace{-0.48em}\textcolor{successpurple}{$\bullet$}\hspace{0.30em}Accuracy}}

\newcommand{\lowgainpanel}[2]{%
  \includegraphics[width=\linewidth,trim=#1,clip]{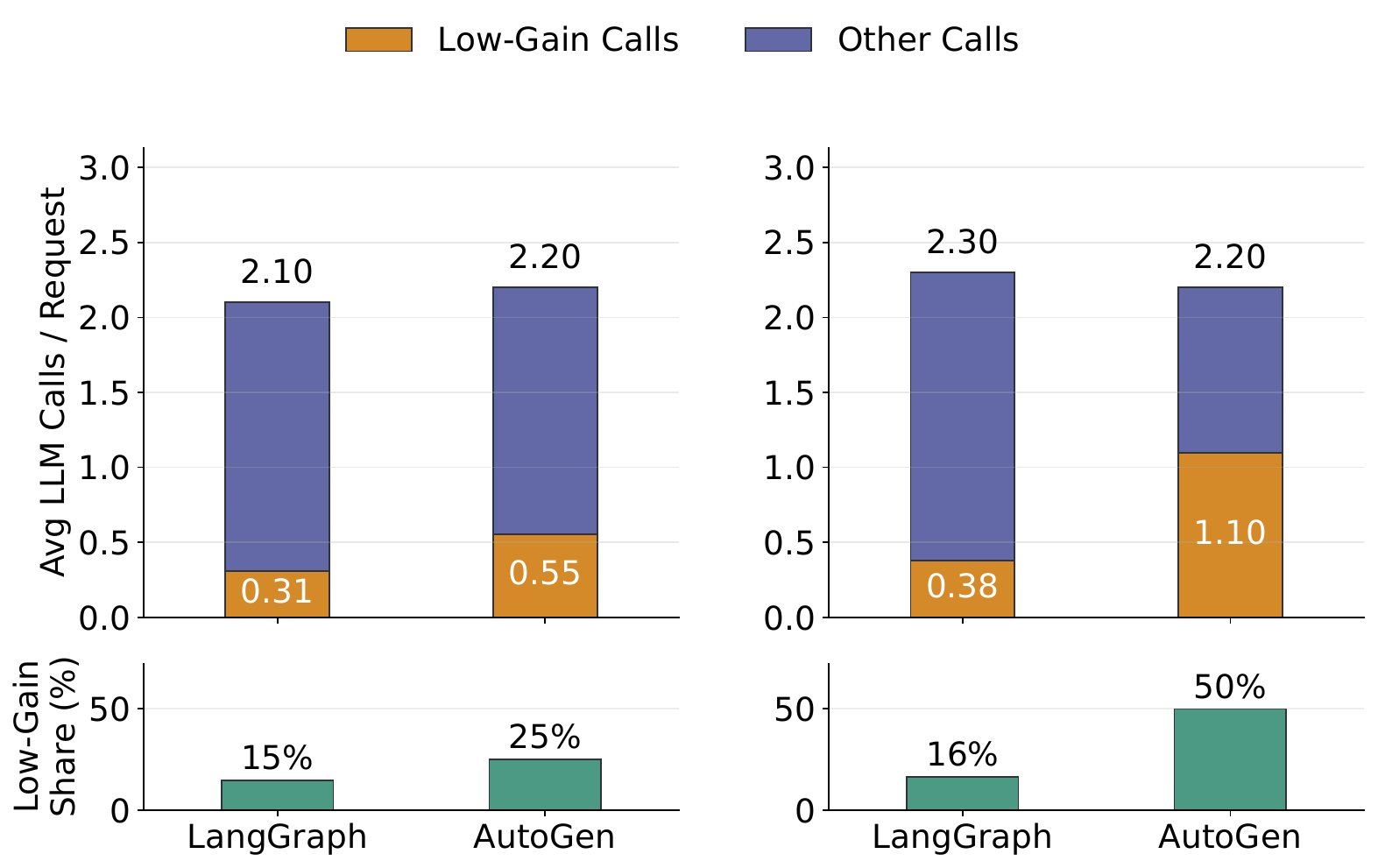}%
  \panelcaption{#2}}
\newcommand{\resultpanel}[4]{%
  \begin{minipage}[b]{#1}
    \centering
    \begin{minipage}[b][#2][b]{\linewidth}
      \centering
      \includegraphics[width=\linewidth,height=#2,keepaspectratio]{#3}
    \end{minipage}
    \panelcaption{#4}
  \end{minipage}}

\newcommand{\trimcaptionpanel}[5]{%
  \begin{minipage}[b]{#1}
    \centering
    \begin{minipage}[b][#2][b]{\linewidth}
      \centering
      \includegraphics[width=\linewidth,height=#2,trim=#5,clip,keepaspectratio]{#3}
    \end{minipage}
    \panelcaption{#4}
  \end{minipage}}
\newcommand{\tokenresultpanel}[5]{%
  \begin{minipage}[b]{#1}
    \centering
      \begin{minipage}[b][#3][b]{#2}
        \centering
        \includegraphics[width=\linewidth,height=#3,trim=0 0 0 83bp,clip,keepaspectratio]{#4}
      \end{minipage}
      \panelcaption{#5}
  \end{minipage}}

\newcommand{\smallqwentokenresultpanel}[5]{%
  \begin{minipage}[b]{#1}
    \centering
      \begin{minipage}[b][#3][b]{#2}
        \centering
        \includegraphics[width=\linewidth,height=#3,trim=0 0 0 90bp,clip,keepaspectratio]{#4}
      \end{minipage}
      \panelcaption{#5}
  \end{minipage}}
\newcommand{\ecdfpanel}[4]{%
  \begin{minipage}[b]{#1}
    \centering
    \begin{minipage}[b][#2][b]{\linewidth}
      \centering
      \includegraphics[width=\linewidth,height=#2,trim=0 0 0 100bp,clip,keepaspectratio]{#3}
    \end{minipage}
    \panelcaption{#4}
  \end{minipage}}
\newcommand{\wideecdfpanel}[4]{%
  \begin{minipage}[b]{#1}
    \centering
    \begin{minipage}[b][#2][b]{\linewidth}
      \centering
      \includegraphics[width=\linewidth,height=#2,trim=0 0 0 100bp,clip,keepaspectratio]{#3}
    \end{minipage}
    \panelcaption{#4}
  \end{minipage}}
\newcommand{\qwenecdfpanel}[4]{%
  \begin{minipage}[b]{#1}
    \centering
    \begin{minipage}[b][#2][b]{\linewidth}
      \centering
      \includegraphics[width=\linewidth,height=#2,trim=0 0 0 100bp,clip,keepaspectratio]{#3}
    \end{minipage}
    \panelcaption{#4}
  \end{minipage}}
\newcommand{\breakdownpanel}[4]{%
  \begin{minipage}[b]{#1}
    \centering
    \begin{minipage}[b][#2][b]{\linewidth}
      \centering
      \includegraphics[width=\linewidth,height=#2,trim=0 0 0 63bp,clip,keepaspectratio]{#3}
    \end{minipage}
    \panelcaption{#4}
  \end{minipage}}
\newcommand{\ftrpanel}[4]{%
  \begin{minipage}[b]{#1}
    \centering
    \begin{minipage}[b][#2][b]{\linewidth}
      \centering
      \includegraphics[width=\linewidth,height=#2,trim=0 0 0 64bp,clip,keepaspectratio]{#3}
    \end{minipage}
    \panelcaption{#4}
  \end{minipage}}
\newcommand{\ftrpanelwide}[4]{%
  \begin{minipage}[b]{#1}
    \centering
    \begin{minipage}[b][#2][b]{\linewidth}
      \centering
      \includegraphics[width=\linewidth,height=#2,trim=0 0 0 64bp,clip,keepaspectratio]{#3}
    \end{minipage}
    \panelcaption{#4}
  \end{minipage}}
\newcommand{\ftrpanelwideb}[4]{%
  \begin{minipage}[b]{#1}
    \centering
    \begin{minipage}[b][#2][b]{\linewidth}
      \centering
      \includegraphics[width=\linewidth,height=#2,trim=0 0 0 64bp,clip,keepaspectratio]{#3}
    \end{minipage}
    \panelcaption{#4}
  \end{minipage}}
\newcommand{\resourcepanel}[4]{%
  \begin{minipage}[b]{#1}
    \centering
    \begin{minipage}[b][#2][b]{\linewidth}
      \centering
      \includegraphics[width=\linewidth,height=#2,keepaspectratio]{#3}
    \end{minipage}
    \panelcaption{#4}
  \end{minipage}}
\newcommand{\qwenresourcepanel}[4]{%
  \begin{minipage}[b]{#1}
    \centering
    \begin{minipage}[b][#2][b]{\linewidth}
      \centering
      \includegraphics[width=\linewidth,height=#2,keepaspectratio]{#3}
    \end{minipage}
    \panelcaption{#4}
  \end{minipage}}
\begin{document}
\bstctlcite{IEEEexample:BSTcontrol}

\title{AgentLoop: Runtime Control of Slot-closed Execution Loops for Tool-augmented LLM Agents}

\author{Wanyi Zheng, Minxian Xu,~\IEEEmembership{Senior Member, IEEE,} Kan Hu, 
Kejiang Ye,~\IEEEmembership{Senior Member, IEEE,} and Chengzhong Xu,~\IEEEmembership{Fellow, IEEE}%
\thanks{This work is supported by National Key R\&D Program of China (No.2026YFE0199800), National Natural Science Foundation of China under Grant 62572462,Guangdong Science and Technology Cooperation Project under Grant	2025A0505020065, Key Research and Development and Technology Transfer Program of Inner Mongolia Autonomous Region (2025YFHH0110) and Shenzhen Science and Technology Program under Grant ZDYJ20251211121533004.}%
\thanks{W. Zheng is with Southern University of Science and Technology, Shenzhen University of Advanced Technology, and Shenzhen Institutes of Advanced Technology, Chinese Academy of Sciences (Email: wy.zheng@siat.ac.cn). M. Xu (corresponding author), K. Hu and K. Ye are with Shenzhen Institutes of Advanced Technology, Chinese Academy of Sciences (Email: mx.xu@siat.ac.cn, kan.hu.1@student.unimelb.edu.au, kj.ye@siat.ac.cn). C. Xu is with Institute of AI and Brain Sciences, Department of Computer Science, University of Macau, Macau, China (Email: czxu@um.edu.mo).}%
}

\markboth{IEEE Transactions on Services Computing}{Zheng \MakeLowercase{\textit{et al.}}: AgentLoop: Runtime Control of Slot-closed Execution Loops for Tool-augmented LLM Agents}

\maketitle

\begin{abstract}
Tool-augmented large language model (LLM) agents are becoming an important execution unit in service computing, but existing agent loops still lack explicit runtime signals for assessing task completion. The challenge lies in the fact that an agent may continue reasoning or invoking services even after the runtime context has stopped changing, while evidence already collected remains unsynthesized into a complete answer, which leads to inefficiency in resource usage. To address these challenges, this paper presents AgentLoop, which provides runtime control of slot-closed execution loops for tool-augmented agents. Slot closure means that the information slots required by a request have been covered by sufficient runtime evidence, and that unresolved slots are explicitly identified before the loop stops. AgentLoop converts open-ended agent iteration into state-driven execution control: it maintains a compact runtime state, uses model-assisted structured verification to check answer completeness and missing evidence, and applies bounded stability and low-gain signals over neighboring LLM/tool rounds before selecting one of three actions: Continue Invocation, Answer Synthesis, or Terminate Iteration. AgentLoop does not change the underlying model or tool set, and instead adds a fine-grained stopping criterion at execution time. Experiments show that AgentLoop reduces redundant execution and context growth, with total token cost reduced by up to 88.44\% and average service invocations reduced by up to 76.85\% against baselines. The ablation study further shows that the slot-centered control path plays a central role, since disabling it increases execution depth and substantially reduces accuracy. Overall, the results suggest that efficient tool-augmented agents can benefit from explicit runtime signals for deciding when further LLM/tool iterations no longer add useful context or supported evidence.
\end{abstract}


\IEEEpeerreviewmaketitle

\section{Introduction}

\IEEEPARstart{L}LM agents extend generation with intermediate reasoning, planning, and external actions. Reasoning methods such as Chain-of-thought, Self-consistency, Tree of Thoughts, and Plan-and-solve organize internal problem solving \cite{wei2022cot,wang2023selfconsistency,yao2023tot,wang2023planandsolve}. Tool-augmented methods such as ReAct, MRKL, and Toolformer connect these states to external services \cite{yao2023react,karpas2022mrkl,schick2023toolformer}. Together, they establish the basic execution pattern of an agent, in which the system interprets a request, invokes services, observes results, and produces an answer.

\begin{figure}[t]
  \centering
  \includegraphics[width=\linewidth]{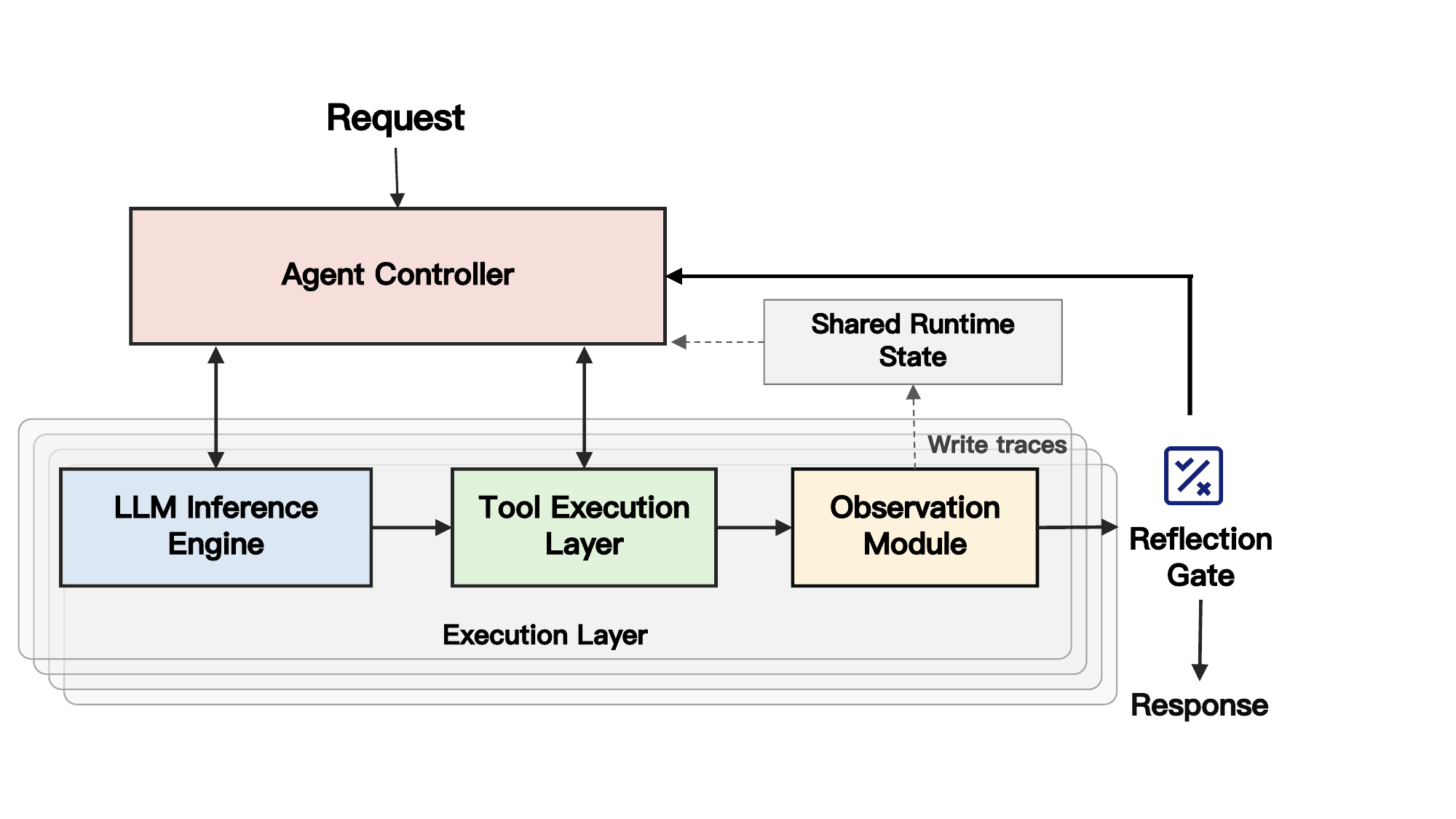}
  \caption{General execution structure of LLM agents.}
  \label{fig:agent_llm_structure}
\end{figure}

Fig.~\ref{fig:agent_llm_structure} illustrates this execution structure. The agent controller does more than issue a single model call. It repeatedly coordinates model inference, tool execution, observation, state writing, and reflection. This structure makes LLM agents flexible, but it also creates a control problem. The system must decide when the shared runtime state contains enough evidence to answer and when another reasoning-action round is still useful.

In service computing, this pattern corresponds to service discovery, service invocation, and result integration. Application programming interface (API)-oriented systems such as Gorilla, ToolLLM, ToolBench, API-Bank, TaskMatrix.AI, and RestGPT improve tool selection, parameter generation, and service composition \cite{patil2024gorilla,qin2024toolllm,li2023apibank,song2023restgpt}. Agent platforms and workflow systems further support open-ended tasks, multi-agent collaboration, and execution optimization \cite{xie2024openagents,wu2023autogen,kim2024llmcompiler}. Related cloud-native and distributed serving work provides the deployment foundation for scalable LLM systems \cite{xu2026cloudnativedistributedsystemsefficient,liao2026dopd}. These advances make tool use increasingly capable, but they also make runtime traces longer and service evidence more diverse.

From the perspective of service computing, LLM agents can be viewed as adaptive service orchestration systems, where foundation models act as reasoning services and external tools represent dynamically invoked computational services. Therefore, efficient agent execution requires not only service selection and composition, but also runtime control mechanisms that determine whether additional service invocations contribute meaningful information  \cite{dadhich2026agenticcontextmanagement}. 

This reinforces that agent reliability depends not only on selecting useful actions, but also on controlling how context accumulates across iterations\cite{li2026acmagenticcontextmanagement}. The remaining systems problem is task-completion awareness. Existing agents commonly stop through a fixed budget, a predefined workflow boundary, or model self-judgment. These mechanisms do not reliably distinguish necessary exploration from repeated LLM reasoning rounds or repeated service calls, and they do not always separate missing evidence from evidence that has already been obtained but not synthesized. Recent evidence on agentic coding shows the scale of this cost. Agentic coding tasks have been shown to consume about 4.17M tokens per task on average, compared with 3.39k tokens for code chat and 1.19k tokens for single-turn code reasoning \cite{bai2026aiagentsspendmoney}. Repeated runs on the same task can vary substantially in total tokens, and higher token use does not necessarily improve accuracy \cite{bai2026aiagentsspendmoney}. Consequently, an agent may continue after the useful context or tool evidence has stabilized, or return tool fragments instead of a complete service result.

This issue affects both service quality and resource use. Each unnecessary round adds model inference, possible tool access, context assembly, and queueing cost, while repeated reasoning or repeated observations may contribute little new information. A practical controller should therefore make two related judgments: whether another LLM-driven reasoning-action round is likely to update the runtime context, and whether another tool invocation is likely to close an information gap or resolve a conflict.

To address the aforementioned challenges, this paper proposes AgentLoop. Unlike existing stopping strategies that rely on predefined budgets or model-generated confidence, AgentLoop performs requirement-aware runtime control by jointly reasoning about information coverage, evidence availability, and marginal utility of future service interactions. AgentLoop uses structured runtime verification to check answer completeness, missing evidence, tool need, confidence, and failure mode, and then applies bounded stability and low-gain\footnote{In this paper, the terms \textit{low-gain} and \textit{low-value} are used interchangeably to describe unnecessary or redundant service invocations that contribute limited additional evidence or context to the runtime state.} signals over neighboring LLM/tool rounds. Based on these signals, AgentLoop suppresses redundant states and selects among Continue Invocation, Answer Synthesis, and Terminate Iteration. AgentLoop is an execution-time layer and does not require changing the base model or tool interfaces.

The key \textbf{contributions} are as follows:
\begin{itemize}
  \item We formulate runtime completion control for tool-augmented LLM agents as a joint decision problem considering requirement coverage, evidence availability, context stability, and execution budget.
  \item We design AgentLoop, a lightweight runtime control architecture consisting of state management, slot verification, redundancy suppression, and iteration control.
  \item We conduct extensive experiments across multiple agent benchmarks and LLM backbones, demonstrating substantial reductions in execution cost while maintaining answer quality.
\end{itemize}

\section{Related Work}

Existing work has improved reasoning, tool use, and orchestration, but task-completion awareness is still rarely treated as a runtime control problem. We group related studies by their main control target and compare them with AgentLoop from the perspective of stopping, context and evidence modeling, and redundant-round control.

\subsection{Reasoning Enhancement and Reflective Generation}

\begin{table*}[t]
  \caption{Capability-based comparison of AgentLoop with representative reasoning, tool-use, and workflow methods.}
  \label{tab:method_comparison}
  \centering
  \scriptsize
  \setlength{\tabcolsep}{3.2pt}
  \renewcommand{\arraystretch}{1.12}
  \begin{tabular}{@{}lccccccccc@{}}
  \toprule
  \multirow{2}{*}{Method} &
  \multicolumn{2}{c}{Task-level State} &
  \multicolumn{3}{c}{Context and Evidence} &
  \multicolumn{3}{c}{Execution Control} \\
  \cmidrule(lr){2-3}\cmidrule(lr){4-6}\cmidrule(l){7-9}
  & \makecell{Req.\\Slots} & \makecell{Answer\\Ready}
  & \makecell{Context\\State} & \makecell{Tool\\Evidence} & \makecell{Gap\\Attribution}
  & \makecell{LLM/Tool\\Utility} & \makecell{Redundancy\\Control} & \makecell{Budget\\Guard} \\
  \midrule
  \makecell[l]{CoT\cite{wei2022cot}, Self-consistency\cite{wang2023selfconsistency}, ToT\cite{yao2023tot}}
    & \dmark & \pmark & \cmark & \dmark & \dmark & \dmark & \dmark & \dmark \\
  \makecell[l]{Reflexion\cite{shinn2023reflexion}, Self-refine\cite{madaan2023selfrefine}}
    & \dmark & \cmark & \cmark & \dmark & \dmark & \pmark & \dmark & \dmark \\
  \makecell[l]{ReAct\cite{yao2023react}, MRKL\cite{karpas2022mrkl}, Toolformer\cite{schick2023toolformer}}
    & \dmark & \pmark & \cmark & \cmark & \dmark & \pmark & \dmark & \dmark \\
  \makecell[l]{Gorilla\cite{patil2024gorilla}, ToolLLM\cite{qin2024toolllm}, API-Bank\cite{li2023apibank}}
    & \dmark & \dmark & \dmark & \cmark & \dmark & \dmark & \dmark & \cmark \\
  \makecell[l]{AutoGen\cite{wu2023autogen}}
    & \dmark & \pmark & \cmark & \pmark & \dmark & \pmark & \cmark & \cmark \\
  \makecell[l]{LangGraph\cite{langchain2026langgraph}}
    & \dmark & \pmark & \cmark & \pmark & \dmark & \pmark & \cmark & \dmark \\
  \textbf{AgentLoop(ours)}
    & \textbf{\cmark} & \textbf{\cmark} & \textbf{\cmark} & \textbf{\cmark}
    & \textbf{\cmark} & \textbf{\cmark} & \textbf{\cmark} & \textbf{\cmark} \\
\bottomrule
  \end{tabular}
  \par\vspace{1mm}
  {\footnotesize \textit{Note:} \cmark means explicit design target, \pmark means partial or indirect support, and \dmark means not a primary part.
  \textit{Req.\ Slots}: the request is decomposed into information items that the final answer must cover.
  \textit{Answer Ready}: whether the current draft already organizes the supported items into a deliverable answer.
  \textit{Context State}: whether runtime context is kept as an explicit state rather than an unstructured trace.
  \textit{Tool Evidence}: whether tool returns are verified as support for the required items, not only stored and replayed.
  \textit{Gap Attribution}: whether an unmet requirement is attributed to missing evidence, to evidence that has been obtained but not yet synthesized, or to an already stable state.
  \textit{LLM/Tool Utility}: whether the marginal value of one more LLM or tool round is estimated before that round is issued.
  \textit{Redundancy Control}: whether rounds that reproduce an already stable state are suppressed.
  \textit{Budget Guard}: whether execution depth, latency, and failure state are guarded as a runtime budget.}
\end{table*}

Chain-of-thought, Self-consistency, Tree of Thoughts, Plan-and-solve, and Plan-and-act improve intermediate reasoning, search, and action organization \cite{wei2022cot,wang2023selfconsistency,yao2023tot,wang2023planandsolve,erdogan2025planact}. Reflexion, Self-refine, and SE-Agent further introduce critique and trajectory revision \cite{shinn2023reflexion,madaan2023selfrefine,guo2025seagent}. These methods help agents reason or revise answers, but their completion signals are still mainly generated by the model itself. They rarely maintain an explicit runtime state that can determine whether another reasoning round changes useful context, whether a missing fact requires another service call, or whether existing evidence only needs to be synthesized. In addition, the stopping rule in these methods is usually either a fixed iteration budget or the model's own judgment that the answer is already good enough, and neither of them separates the case where the context has stopped changing from the case where the task still has an unfilled requirement.

\subsection{Tool-augmented Language Models}

ReAct\cite{yao2023react}, MRKL\cite{karpas2022mrkl}, Toolformer\cite{schick2023toolformer}, Gorilla\cite{patil2024gorilla}, AvaTaR\cite{wu2024avatar}, ToolLLM\cite{qin2024toolllm}, ToolBench, API-Bank\cite{li2023apibank}, ART\cite{paranjape2023art}, Chameleon\cite{lu2023chameleon}, and LLMCompiler\cite{kim2024llmcompiler} improve tool selection, API grounding, composition, or parallel execution. Tool-ecosystem studies further optimize tool construction and instructions \cite{wolflein2025toolmaker,wu2025toolutil}. Their main objective is to make tool access more accurate and scalable. However, successful tool access is not equivalent to task completion: these systems usually store tool returns or traces, but do not explicitly check whether another LLM/tool iteration would update useful context, add new supported evidence, or repeat an already stable state. In practice this appears as continued invocation on top of an already stable tool result, where later calls return the same content and add only latency and context length.

\subsection{Agent Evaluation and Workflow Orchestration}

AgentBench\cite{liu2024agentbench} and GAIA\cite{mialon2023gaiabenchmarkgeneralai} expose agent capability and grounded-task failure modes. Conversational Information Gain (CIG) measures how utterances advance deliberative dialogues by updating a semantic memory state, providing a related but evaluation-oriented view of information gain \cite{chen-etal-2026-cig}. AutoGen, AFlow, ADAS, GPTSwarm, Agent Workflow Memory, evolving orchestration, and Agentix improve conversation, workflow construction, graph search, memory, or serving \cite{wu2023autogen,zhang2025aflow,hu2025adas,zhuge2024gptswarm,wang2025awm,dang2025orchestration,luo2026agentix,bai2026oracl}. These systems constrain or optimize execution at the platform level. Their stopping conditions are often tied to nodes, branches, workflow boundaries, or budgets, while an open tool node may still lack a local rule for distinguishing unchanged LLM context, missing evidence, and evidence that has been collected but not yet synthesized.

Table~\ref{tab:method_comparison} summarizes this distinction as a capability matrix. The table is meant to show that existing methods usually cover only part of the control stack, while AgentLoop combines request-level state representation with evidence-aware control of both LLM and tool iterations. For instance, AutoGen and LangGraph do not  decompose requests into items that the final answer to cover (request slot), and do not consider the missing evidence can lead to unmet requirement (gap attribution). 

The distinction is therefore not that AgentLoop replaces reasoning, tool-use, or workflow frameworks. It adds a runtime control layer that can be embedded into these systems, closes request-level information slots, distinguishes missing evidence from unsynthesized evidence, and suppresses low-gain LLM or tool iterations before they expand the execution trace.

\begin{figure*}[!htb]
	\centering
	\makebox[\textwidth][c]{%
		\resultpanel{0.33\textwidth}{0.12\textheight}{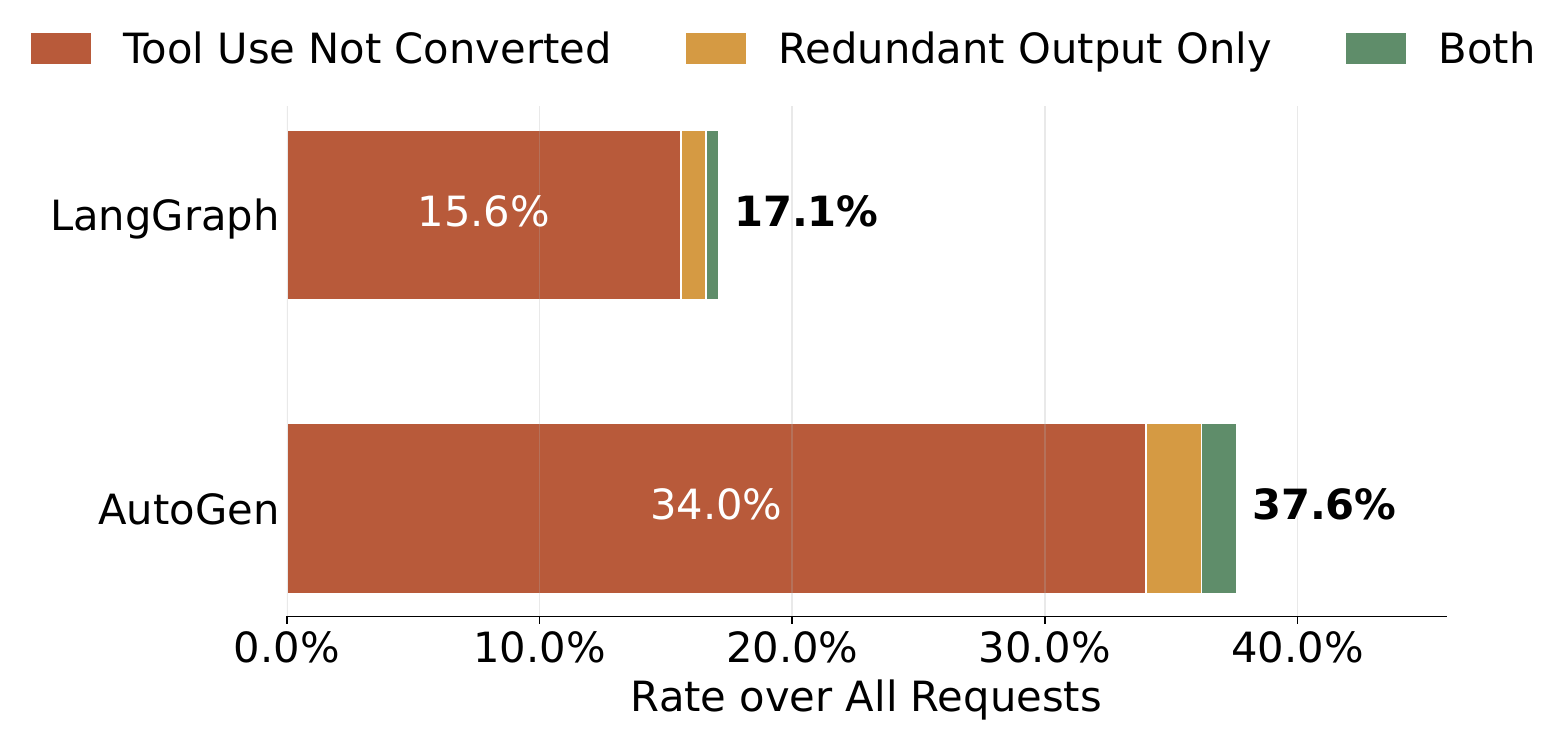}{(a) Overall risk.}%
		\hspace{0.012\textwidth}%
		\resultpanel{0.31\textwidth}{0.12\textheight}{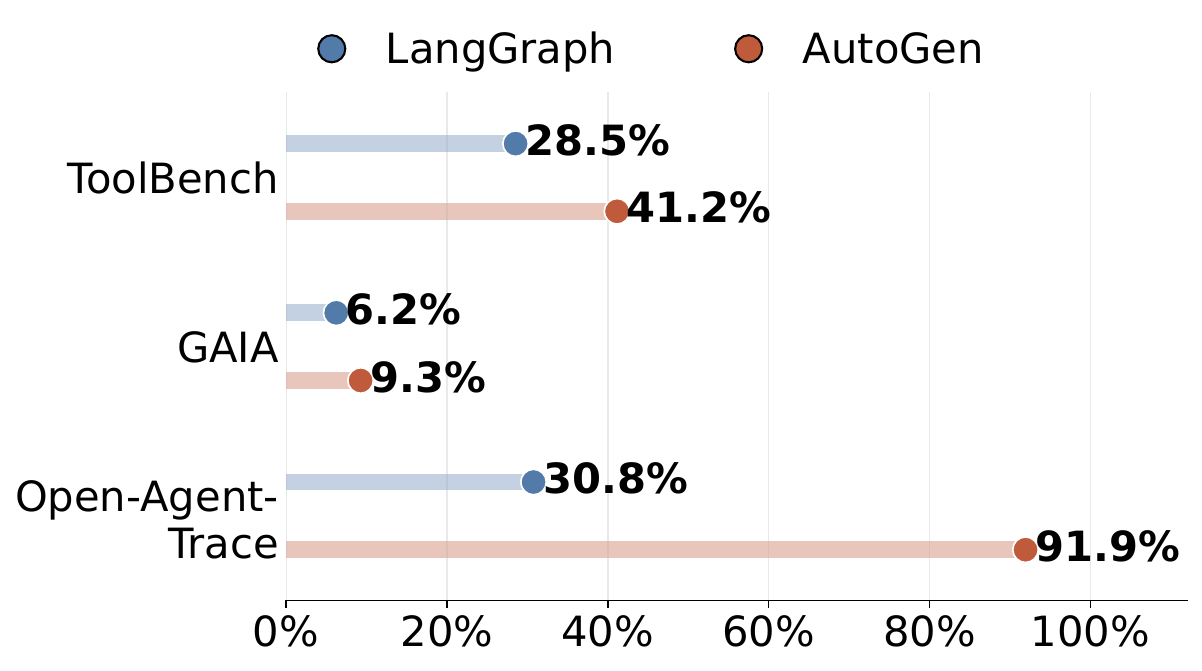}{(b) Dataset risk.}%
		\hspace{0.012\textwidth}%
		\trimcaptionpanel{0.33\textwidth}{0.14\textheight}{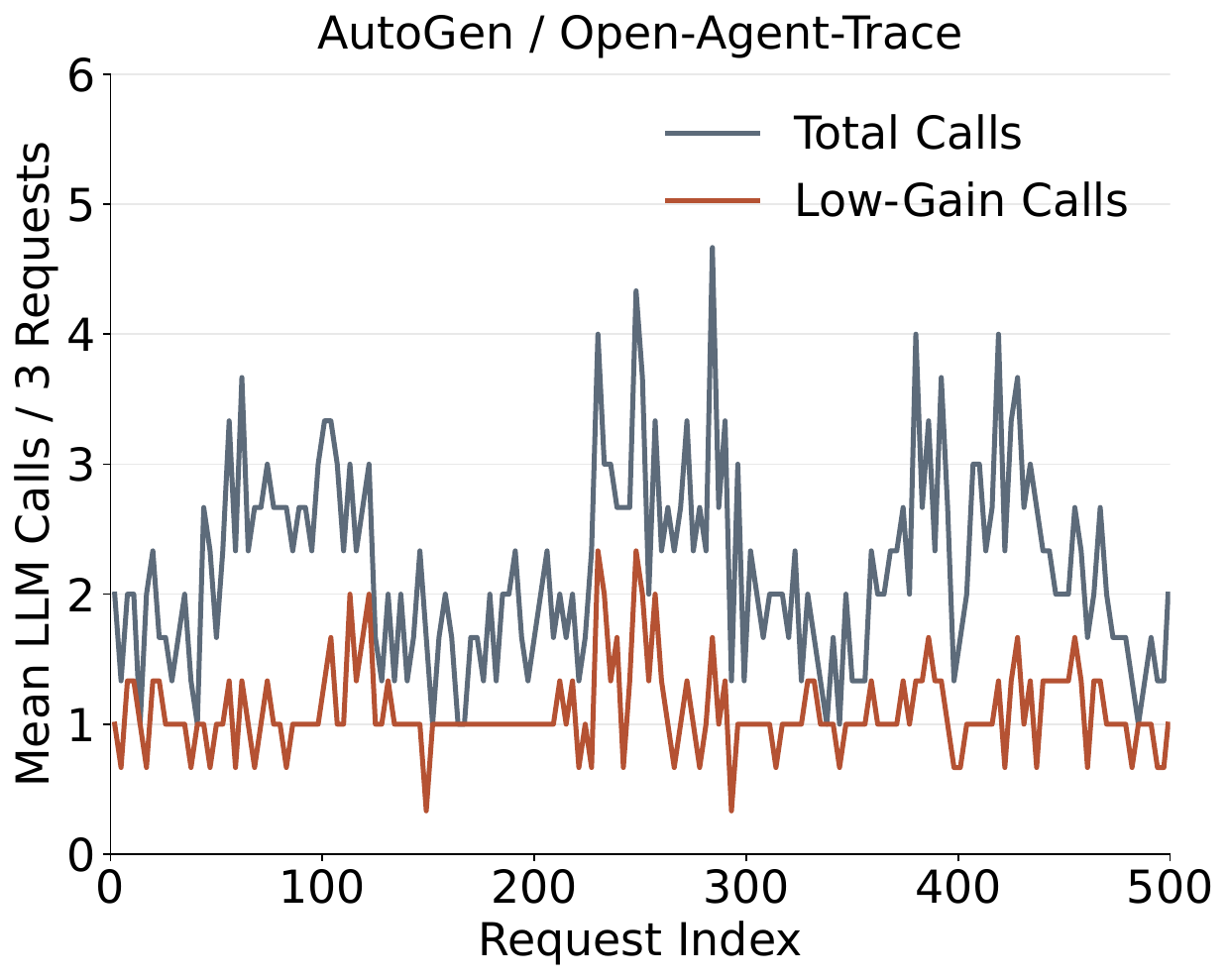}{(c) LLM-call trend.}{0 5bp 0 0}%
	}
	\caption{Motivation overview.}
	\label{fig:motivation_overview}
\end{figure*}

\section{Problem Motivation}

We summarize the need for runtime completion control through three observations. We use \emph{low-value execution risk} here as a request-level operational indicator for repetitive or low-yield execution: the agent has already invoked tools or produced repeated LLM outputs, but the final state still fails to add useful supported information. Operationally, this indicator covers post-tool failure and repeated model outputs in the recorded trace, and it functions as a direct control signal for repetitive runtime behavior rather than a learned utility estimate. This motivation study uses the same unified GPU-node and runtime configuration as the main experiments, with cache disabled and the same request order within each backbone.

\paragraph{Observation 1}
\emph{Tool invocation by itself does not guarantee that request-level evidence is complete.}
Fig.~\ref{fig:motivation_overview}(a) shows that the low-value execution pattern is clearly visible in both workflow-based and conversation-based baselines. The overall risk reaches 17.1\% for LangGraph and 37.6\% for AutoGen, with post-tool failure instances making up most of the count. Fig.~\ref{fig:motivation_overview}(b) further shows that the risk rises on ToolBench and Open-Agent-Trace, where service composition and evidence integration are more demanding. The central issue is therefore not only whether an agent can invoke a service, but whether it can recognize, verify, and synthesize the returned evidence.

\paragraph{Observation 2}
\emph{Low-gain LLM behavior is a persistent runtime pattern rather than an isolated failure case.}
Fig.~\ref{fig:motivation_overview}(c) shows that low-gain LLM calls remain close to total LLM calls across the Open-Agent-Trace request sequence. The low-gain share reaches 49.82\% under request-level aggregation and 53.08\% when three-request windows are averaged. Despite the different averaging orders, both measurements indicate that low-gain execution is a recurring runtime state. This motivates online monitoring at the agent-iteration level, so AgentLoop should detect both LLM rounds that no longer update useful context and tool invocations that no longer add supported evidence.

\begin{figure}[h]
	\centering
	\includegraphics[width=0.56\linewidth,trim=150bp 395bp 150bp 0bp,clip]{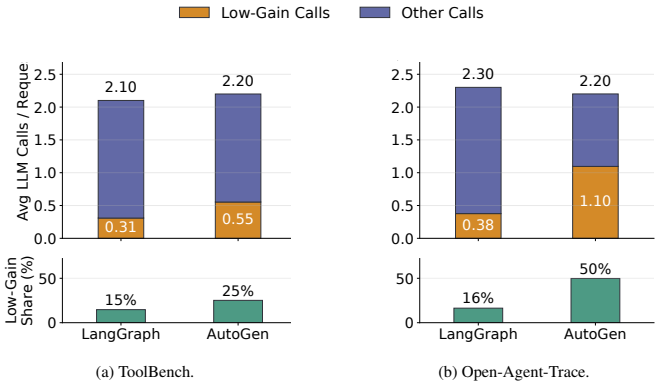}
	
	\vspace{0.2mm}
	
	\begin{minipage}[b]{0.435\linewidth}
		\centering
		\lowgainpanel{0 6bp 390bp 118bp}{(a) ToolBench.}
	\end{minipage}
	\hfill
	\begin{minipage}[b]{0.435\linewidth}
		\centering
		\lowgainpanel{390bp 6bp 0 118bp}{(b) Open-Agent-Trace.}
	\end{minipage}
	\caption{LLM-call composition on ToolBench and Open-Agent-Trace.}
	\label{fig:motivation_llm_composition}
\end{figure}

\paragraph{Observation 3}
\emph{Similar LLM-call depth can hide very different ratios of productive and low-gain execution.}
Fig.~\ref{fig:motivation_llm_composition} shows that LangGraph and AutoGen have similar average LLM-call depth on ToolBench and Open-Agent-Trace, yet their low-gain components differ substantially. On Open-Agent-Trace, AutoGen has a 49.82\% aggregate low-gain share compared with 16.38\% for LangGraph. Thus, LLM-call count alone is insufficient for evaluating execution efficiency, and the controller must also judge whether each reasoning-action round changes the useful runtime context and whether associated tool calls add supported evidence.

\section{AgentLoop Design}

Based on the aforementioned observations, to address incomplete evidence integration and repeated low-gain execution, we design AgentLoop as a runtime slot-closed control layer for tool-augmented LLM agents. It represents request requirements as slots and uses runtime context and tool evidence to decide whether to continue invocation, synthesize an answer, or terminate the iteration.

\subsection{System Overview}

AgentLoop adds a runtime control layer around an existing tool-augmented agent. Given the request, available tools, current answer draft, and execution trace, it chooses \emph{Continue Invocation}, \emph{Answer Synthesis}, or \emph{Terminate Iteration}, while leaving the base model and external tools unchanged.

\begin{figure}[t]
  \centering
  \includegraphics[width=\linewidth]{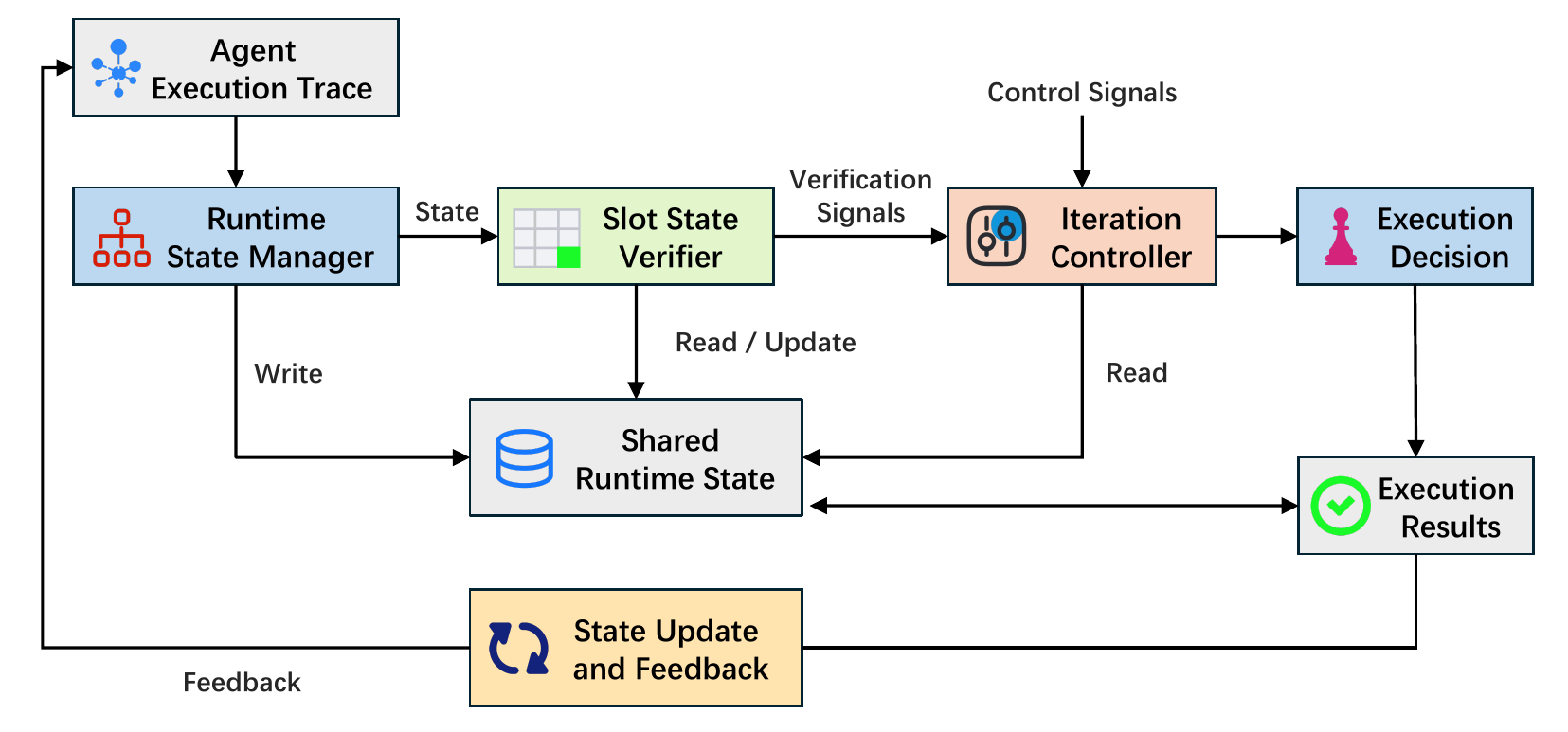}
  \caption{Method schematic of AgentLoop.}
  \label{fig:method_schematic}
\end{figure}

\begin{figure*}[t]
  \centering
  \includegraphics[width=0.96\textwidth]{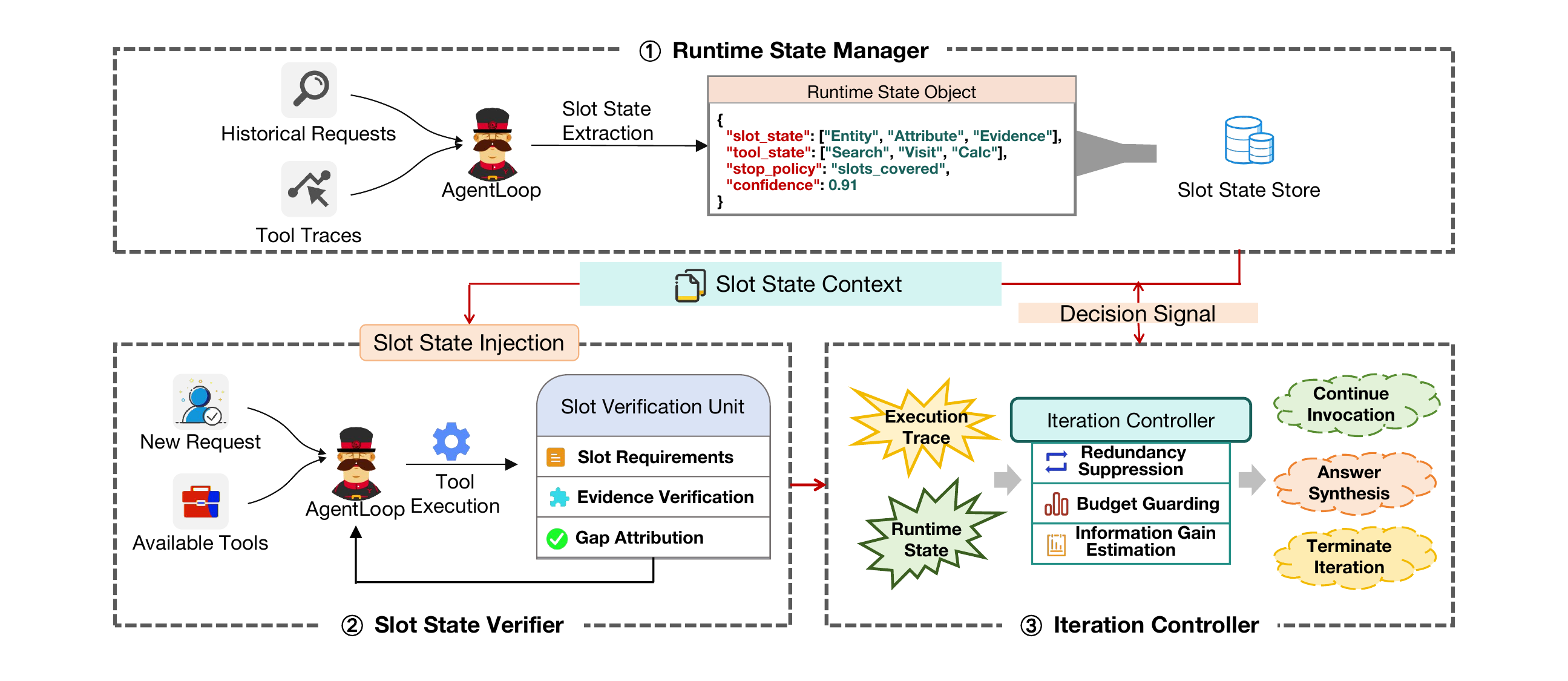}
  \caption{Overview of the AgentLoop runtime control architecture.}
  \label{fig:system_architecture}
\end{figure*}

Fig.~\ref{fig:method_schematic} summarizes the state-mediated data flow: the execution trace is compressed into shared runtime state, checked by the Slot State Verifier, and consumed by the Iteration Controller to produce the next execution decision. Given only the request, the current answer draft, and the compressed evidence summary, the verifier performs three checks: 1) whether the answer draft already covers the information that the request requires; 2) whether the covered requirements are actually supported by the evidence obtained so far, rather than asserted by the model; and 3) if a requirement is still unmet, whether the gap is attributable to evidence that has not been retrieved, to evidence that has been retrieved but not yet organized into the answer, or to a tool state that is already stable or self-contradictory. The resulting execution feedback updates the shared state before the next agent round, and the detailed component responsibilities are shown in Fig.~\ref{fig:system_architecture}.

The control principle is that AgentLoop separates three cases that are often conflated: 1) missing evidence may require another tool invocation; 2) stable evidence with an incomplete response requires Answer Synthesis; and 3) stable context with sufficient, supported evidence permits Terminate Iteration. Redundancy Suppression is therefore a decision signal over both LLM rounds and tool calls rather than a rule that forbids repeated use of a tool.

Fig.~\ref{fig:system_architecture} shows the closed-loop architecture and its three main components. The \emph{Runtime State Manager} extracts slot state from historical requests and tool traces, writes a compact Runtime State Object into the Slot State Store, and exposes the resulting Slot State Context to the online loop. In the implementation, this runtime state is a compact summary rather than a full trace dump: it keeps the latest request context, the latest tool evidence, and the control flags needed by the next round. Using the request, available tools, and slot context, the \emph{Slot State Verifier} checks Slot Requirements, Evidence Verification, and Gap Attribution. The \emph{Iteration Controller} reads the execution trace and runtime state, applies Redundancy Suppression, Budget Guarding, and a low-gain stability heuristic, and selects the next execution decision based on LLM-context changes and new tool evidence.

\section{System Modeling and Algorithm Design}
In this section, we introduce the system modeling and algorithm design in AgentLoop. 
\subsection{Online Runtime Mechanism}

After runtime stage $t$, AgentLoop represents the state observed by its controller as
\begin{equation}
X_t=(q,Y_t,\mathit{Call}_t,\bar{E}_t,V_t,H_t,B_t).
\label{eq:runtime_state}
\end{equation}
Here, $q$ is the user request, $Y_t$ is the current answer draft, and $\mathit{Call}_t$ is the set of tool calls issued in the current or most recent round. $\bar{E}_t$ is the compressed summary of tool states and returned evidence. Operationally, it keeps only a bounded textual representation of each new tool result, such as the tool name, error flag, key finding, and unresolved fields, rather than the full raw trace. $V_t$ is the structured verifier output, $H_t$ is the history of dialogue messages, tool calls, tool results, and control events, and $B_t$ contains runtime boundaries and configuration such as iteration limits, reflection limits, timeout constraints, similarity thresholds, and feature switches. Equation~\eqref{eq:runtime_state} is a state abstraction rather than a separately instantiated Python class. It corresponds to the implementation fields that maintain model context, current model results, previous calls, previous execution results, and the tool-state cache.

The verifier reads the request, current answer, and compressed evidence state and returns
\begin{equation}
V_t=\operatorname{Verify}(q,Y_t,\bar{E}_t)=(a_t,M_t,n_t,\kappa_t,f_t).
\label{eq:verifier}
\end{equation}
The binary variable $a_t$ is the \texttt{answer\_complete} flag. $M_t$ is the list of missing evidence or unresolved requirements, and $n_t$ is the \texttt{need\_tool} flag. The confidence value $\kappa_t\in[0,1]$ is the verifier confidence. The failure state $f_t$ records conditions such as \texttt{ok}, \texttt{answer\_incomplete}, \texttt{tool\_set\_unstable}, \texttt{tool\_conflict}, and \texttt{low\_confidence}. Together, $M_t$, $n_t$, and $f_t$ implement Gap Attribution: an unmet requirement is attributed to missing evidence when $n_t=1$, to evidence that has already been obtained but not yet synthesized when $n_t=0$ and $f_t=\texttt{answer\_incomplete}$, and to an already stable or contradictory tool state when $f_t$ is \texttt{tool\_set\_unstable} or \texttt{tool\_conflict}. In other words, the slot state is a fixed-schema verifier record generated from the current request, the current answer, and the compressed evidence summary, rather than a rule-based slot extractor. Historical requests and tool traces are retained in $H_t$ and $\bar{E}_t$ as execution context, but the verifier itself consumes only the current round state. When evidence is missing or contradictory, the failure labels guide the controller toward another tool round or answer integration. The verifier is a structured auxiliary-model output used by the controller. It is not an independent module named \texttt{SlotGen}.

The online closure test is therefore binary:
\begin{equation}
C_t=\mathbb{I}[a_t=1\land n_t=0].
\label{eq:closure}
\end{equation}
The indicator $\mathbb{I}[\cdot]$ takes value one when its condition is true and zero otherwise. Thus, $C_t=1$ means that the verifier regards the answer as complete and does not request another tool round. It is a direct closure signal, not just a weighted coverage score.

To compare adjacent tool-call plans, AgentLoop uses the mixed text similarity and the resulting call-state similarity in one definition:
\begin{equation}
\resizebox{0.98\columnwidth}{!}{$
\begin{aligned}
\rho(x,y)&=0.55\operatorname{Lex}(x,y)+0.45\operatorname{Jac}(\operatorname{Token}(x),\operatorname{Token}(y)),\\
S^{\mathrm{call}}_t&=
\begin{cases}
1, & \mathit{Call}_{t-1}=\mathit{Call}_t=\emptyset,\\
0, & \mathit{NM}_t=0\ \lor\ \mathit{XOR}_t=1,\\
\displaystyle\frac{1}{m}\sum_{j=1}^{m}\rho(\mathit{arg}_{t-1,j},\mathit{arg}_{t,j}), & \text{otherwise}.
\end{cases}
\end{aligned}
$}
\label{eq:call_similarity}
\end{equation}
Here, $\operatorname{Lex}$ is normalized lexical similarity, $\operatorname{Token}$ returns the token set of a text, and $\operatorname{Jac}$ is Jaccard similarity. $\mathit{NM}_t$ is one when the tool-name multisets match, and $\mathit{XOR}_t$ is one when exactly one call set is empty. The parameter text $\mathit{arg}_{t,j}$ is the normalized argument of the $j$th aligned call and $m$ is the number of aligned calls. The implementation groups calls by tool name and sorts their normalized JSON arguments before comparison. The mixed score is a lexical and token-set comparison, not an embedding similarity.

The corresponding result-state similarity is where $\mathit{Result}_t$ is the set of tool results in round $t$, $\mathit{res}_{t,j}$ is an aligned normalized result text, and $k$ is the number of aligned results. $\mathit{Mismatch}_t$ is one when either result set is empty, tool grouping or multiplicity differs, or an error state differs. The predicate $\operatorname{is\_error}$ records whether a result represents a tool error. The comparison checks these fields before comparing compressed result text.
\begin{equation}
\resizebox{0.98\columnwidth}{!}{$
S^{\mathrm{result}}_t=
\begin{cases}
1, & \mathit{Result}_{t-1}=\mathit{Result}_t=\emptyset,\\
0, & \mathit{Mismatch}_t=1,\\
\displaystyle\frac{1}{k}\sum_{j=1}^{k}\rho(\mathit{res}_{t-1,j},\mathit{res}_{t,j}), & \text{otherwise}.
\end{cases}
 $}
\label{eq:result_similarity}
\end{equation}

Let $\operatorname{NameMatch}_t$ indicate equality of tool-name multisets. Let $\operatorname{ExactCall}_t$ and $\operatorname{ExactResult}_t$ indicate equality of the normalized call and result signatures. The stability gate is
\begin{equation}
\resizebox{0.98\columnwidth}{!}{$
\begin{aligned}
G^{\mathrm{stab}}_t=\mathbb{I}\!\bigl[&\mathit{NM}_t=1\\
&\land(\operatorname{ExactCall}_t=1\lor
\operatorname{EnableCallSim}=1\land S^{\mathrm{call}}_t\ge\tau_{\mathrm{call}})\\
&\land(\operatorname{ExactResult}_t=1\lor
\operatorname{EnableResultSim}=1\land S^{\mathrm{result}}_t\ge\tau_{\mathrm{result}})\bigr].
\end{aligned}
 $}
\label{eq:stability_gate}
\end{equation}
Here, $\mathit{NM}_t$ is the name-match variable defined above. $\operatorname{ExactCall}_t$ and $\operatorname{ExactResult}_t$ indicate equality of normalized call and result signatures. $\operatorname{EnableCallSim}$ and $\operatorname{EnableResultSim}$ are configuration switches. The default thresholds are fixed runtime settings rather than theoretical constants. The gate means that the neighboring action and result states are stable enough that further replanning has low expected value.

The implementation uses discrete branches rather than a continuous utility optimizer. A unified abstraction of these branches is
\begin{equation}
\resizebox{0.98\columnwidth}{!}{$
\pi_t=
\begin{cases}
\texttt{need\_tool\_round}, & n_t=1\lor M_t\neq\emptyset,\\
\texttt{verifier\_retry}, & f_t=\texttt{tool\_conflict},\\
\texttt{rollback\_retry}, & f_t=\texttt{tool\_set\_unstable},\\
\texttt{rollback\_retry}, & f_t=\texttt{low\_confidence}\land\kappa_t<\tau_{\mathrm{retry}},\\
\texttt{integrate\_answer\_retry\_break}, & f_t=\texttt{answer\_incomplete}\land n_t=0\land M_t=\emptyset,\\
\texttt{stop\_replanning}, & C_t=1\land G^{\mathrm{stab}}_t=1\land f_t=\texttt{ok},\\
\texttt{skip\_decode\_output\_break}, & C_t=1\land \kappa_t\ge\tau_{\mathrm{gate}},\\
\texttt{skip\_decode\_output}, & \kappa_t\ge\tau_{\mathrm{gate}},\\
\texttt{keep\_decode}, & \text{otherwise}.
\end{cases}
 $}
\label{eq:control_policy}
\end{equation}
Here, $\pi_t$ is a mathematical abstraction of the code's control branches. It is not the name of a standalone function in the implementation. The branches are evaluated from top to bottom. Unresolved evidence, conflicting tool states, and unstable tool states take priority over stability-based stopping. The thresholds $\tau_{\mathrm{retry}}$ and $\tau_{\mathrm{gate}}$ are fixed runtime settings for low-confidence retry and decode skipping. In particular, $G^{\mathrm{stab}}_t=1$ only signals that the neighboring tool state is stable. It triggers \texttt{stop\_replanning} only when the closure signal is active and the verifier state is \texttt{ok}. In Algorithm~\ref{alg:agentloop}, the branches of Equation~\eqref{eq:control_policy} are grouped by whether the bounded loop stays active: branches that request another model or tool round map to $\mathsf{CI}$, the answer-integration branch maps to $\mathsf{AS}$, and the closure branches map to $\mathsf{TI}$.

The state in Fig.~\ref{fig:system_architecture} is compressed by the Runtime State Manager, checked by the structured verifier, and consumed by the Iteration Controller. If evidence is missing, $n_t=1$ can request another tool round. If evidence is already available but $a_t=0$, answer integration is preferred. If the action and result signatures are stable and the answer state is closed, the controller stops replanning. This separation prevents offline diagnostic metrics from being mistaken for online control inputs.

\begin{algorithm}[!t]
	\caption{AgentLoop Closed-loop Execution Control}
	\label{alg:agentloop}
	\footnotesize
	\setlength{\jot}{1pt}
	\begin{algorithmic}[1]
		\RaggedRight
		\REQUIRE $q$: request; $\mathcal{T}$: tool set;
		$B=(T_{\max},\Theta,\Omega)$: runtime budget
		\ENSURE $Y^*$: final answer
		\STATE $\begin{aligned}[t]
			&T_{\max}:\text{ maximum number of rounds};\quad
			\Theta:\text{ thresholds};\\[-0.5mm]
			&\Omega:\text{ similarity switches};\quad
			\mathcal{A}=\{\mathsf{CI},\mathsf{AS},\mathsf{TI}\}
		\end{aligned}$
		\STATE $\begin{aligned}[t]
			&\mathsf{CI}:=\text{Continue Invocation};\quad
			\mathsf{AS}:=\text{Answer Synthesis};\\[-0.5mm]
			&\mathsf{TI}:=\text{Terminate Iteration}
		\end{aligned}$
		\STATE $\begin{aligned}[t]
			&Y:\text{ answer draft};\quad
			\bar{E}:\text{ evidence summary};\\[-0.5mm]
			&H:\text{ execution history};\quad
			\mathit{Call}_t,\mathit{Result}_t:
			\text{ round states}
		\end{aligned}$
		\STATE $\begin{aligned}[t]
			&V_t=(a_t,M_t,n_t,\kappa_t,f_t):
			\text{ verifier state};\\[-0.5mm]
			&a_t,n_t\in\{0,1\};\quad
			\kappa_t\in[0,1];\\[-0.5mm]
			&M_t:\text{ missing set};\quad
			f_t:\text{ failure state}
		\end{aligned}$
		\STATE $\begin{aligned}[t]
			&C_t:\text{ closure};\quad
			S_t^{\mathrm{call}},S_t^{\mathrm{result}}:
			\text{ similarities};\\[-0.5mm]
			&G_t^{\mathrm{stab}}:\text{ stability gate};\quad
			\pi_t\in\mathcal{A}:\text{ control action}
		\end{aligned}$
		\STATE $\begin{aligned}[t]
			&\mathbb{I}[\cdot]\in\{0,1\};\\[-0.5mm]
			&\operatorname{SynthesizeEvidence}(Y,\bar{E}):
			\text{ answer synthesis};\\[-0.5mm]
			&\operatorname{ReturnGroundedAnswer}(Y,\bar{E}):
			\text{ grounded answer}
		\end{aligned}$
		\STATE \textbf{\textit{// Initialization}}
		\STATE $\begin{aligned}[t]
			&(Y,\mathit{Call}_{-1},\mathit{Result}_{-1})
			\leftarrow
			(\varnothing,\varnothing,\varnothing);\\[-0.5mm]
			&(\bar{E},H,t)
			\leftarrow
			(\varnothing,\varnothing,0)
		\end{aligned}$
		\STATE \textbf{\textit{// Online execution and state update}}
		\WHILE{$t<T_{\max}$}
		\STATE $\begin{aligned}[t]
			&(Y,\mathit{Call}_t)
			\leftarrow
			\operatorname{GenerateReasoningCalls}\\[-0.5mm]
			&\quad(q,Y,\bar{E},H)
		\end{aligned}$
		\STATE $\begin{aligned}[t]
			&(H,\mathit{Result}_t,\bar{E})
			\leftarrow
			\operatorname{ExecuteToolsUpdateState}\\[-0.5mm]
			&\quad(\mathcal{T},\mathit{Call}_t,H)
		\end{aligned}$
		\STATE $\begin{aligned}[t]
			&V_t
			\leftarrow
			\operatorname{VerifyAnswerState}(q,Y,\bar{E})\\[-0.5mm]
			&\quad=(a_t,M_t,n_t,\kappa_t,f_t)
		\end{aligned}$
		\STATE $\begin{aligned}[t]
			&C_t
			\leftarrow
			\mathbb{I}\!\left[
			a_t=1\land n_t=0
			\right]
		\end{aligned}$
		\STATE $\begin{aligned}[t]
			&(S_t^{\mathrm{call}},
			S_t^{\mathrm{result}},
			G_t^{\mathrm{stab}})
			\leftarrow\\[-0.5mm]
			&\quad
			\operatorname{ComputeStateStability}
			\bigl(
			V_t,
			\mathit{Call}_{t-1},
			\mathit{Call}_t,\\[-0.5mm]
			&\qquad
			\mathit{Result}_{t-1},
			\mathit{Result}_t
			\bigr)
		\end{aligned}$
		\STATE
		\textbf{\textit{// Control decision and action dispatch}}
		\STATE $\begin{aligned}[t]
			&\pi_t
			\leftarrow
			\operatorname{SelectControlAction}\\[-0.5mm]
			&\quad(C_t,V_t,G_t^{\mathrm{stab}},B)
		\end{aligned}$
		\IF{$\pi_t=\mathsf{CI}$}
		\STATE $\begin{aligned}[t]
			&(\mathit{Call}_{t-1},
			\mathit{Result}_{t-1},t)
			\leftarrow\\[-0.5mm]
			&\quad
			(\mathit{Call}_t,
			\mathit{Result}_t,t+1)
		\end{aligned}$
		\ELSIF{$\pi_t=\mathsf{AS}$}
		\STATE $\begin{aligned}[t]
			&Y^*
			\leftarrow
			\operatorname{SynthesizeEvidence}(Y,\bar{E})
		\end{aligned}$
		\RETURN $Y^*$
		\ELSIF{$\pi_t=\mathsf{TI}$}
		\STATE $\begin{aligned}[t]
			&Y^*
			\leftarrow
			\operatorname{ReturnGroundedAnswer}(Y,\bar{E})
		\end{aligned}$
		\RETURN $Y^*$
		\ENDIF
		\ENDWHILE
		\STATE $\begin{aligned}[t]
			&Y^*
			\leftarrow
			\operatorname{ReturnGroundedAnswer}(Y,\bar{E})
		\end{aligned}$
		\RETURN $Y^*$
	\end{algorithmic}
\end{algorithm}

\subsection{Trace Diagnostics and Algorithms}

The next four equations define offline diagnostics from completed request traces. Let $u_i$ be the number of tool calls for request $i$, and let $c_i\in\{0,1\}$ be its final correctness flag, where one means correct and zero means incorrect. The post-tool failure flag is
\begin{equation}
p_i=\mathbb{I}[u_i>0\land c_i=0].
\label{eq:post_tool_failure}
\end{equation}
This flag does not measure the tool error rate. It marks a request that used at least one tool but still ended incorrectly.

Let $\operatorname{inter\_repeat}_i$ indicate repeated output across adjacent answer rounds and let $\operatorname{intra\_repeat}_i$ indicate duplicated content within the final answer. The repetition-risk flag is
\begin{equation}
r_i=\mathbb{I}[\operatorname{inter\_repeat}_i=1\lor\operatorname{intra\_repeat}_i=1].
\label{eq:repeat_risk}
\end{equation}
The combined low-value execution marker is
\begin{equation}
z_i=\mathbb{I}[p_i=1\lor r_i=1].
\label{eq:low_gain_flag}
\end{equation}
This is a trace-based operational definition built for execution analysis. It is not a Shannon entropy, KL divergence, or strict information-gain estimate, and it serves as a practical control indicator for repetitive or low-yield execution.

For $N$ requests, the low-gain request rate (LGR) is
\begin{equation}
\operatorname{LGR}=\frac{\sum_{i=1}^{N}z_i}{N}.
\label{eq:lgr}
\end{equation}
The low-gain tool-call share (LGTS) is
\begin{equation}
\operatorname{LGTS}=\begin{cases}
\displaystyle\frac{\sum_{i=1}^{N}u_i z_i}{\sum_{i=1}^{N}u_i}, & \sum_{i=1}^{N}u_i>0,\\
0, & \text{otherwise}.
\end{cases}
\label{eq:lgts}
\end{equation}
Here, LGR is request-level, while LGTS attributes tool calls to flagged requests. Both are computed after the trace is complete and are used for diagnosis rather than online stopping. In that sense, they summarize workload-level risk patterns instead of estimating the marginal value of any individual tool call.

Algorithm~\ref{alg:agentloop} describes the online closed-loop execution procedure. Lines 1--8 define the runtime interface, control parameters, action abbreviations, state variables, and initialization. The online execution stage performs one model/tool round, updates the compressed evidence, verifies answer completeness, and compares neighboring call/result states to obtain the control signals (lines 9--15). The controller then selects and dispatches one of the three actions: continue invocation, synthesize the available evidence, or terminate with an evidence-grounded answer (lines 16--25). The bounded loop closes at line 26, and lines 27--28 provide a fallback answer when the round limit is reached. Thus, the algorithm separates evidence acquisition from answer synthesis and avoids extending the loop when further execution is unlikely to add useful information.

\begin{algorithm}[!t]
\caption{Low-value Execution Risk Detection}
\label{alg:low_gain}
\footnotesize
\begin{algorithmic}[1]
\REQUIRE $T=\{(u_i,c_i,\operatorname{inter\_repeat}_i,\operatorname{intra\_repeat}_i)\}_{i=1}^{N}$: completed traces
\ENSURE $\{z_i\}_{i=1}^{N}$, $\operatorname{LGR}=n_g/N$, $\operatorname{LGTS}=u_g/\max(1,u)$
\STATE $u_i\in\mathbb{N}_{0}$; $c_i,\operatorname{inter\_repeat}_i,\operatorname{intra\_repeat}_i\in\{0,1\}$; $\mathbb{I}[\cdot]\in\{0,1\}$
\STATE $n_g$: flagged requests; $u_g$: flagged calls; $u$: total calls
\STATE $(n_g,u_g,u)\leftarrow(0,0,0)$
\STATE \textbf{\texttt{// Risk detection}}
\FOR{$i=1,\ldots,N$}
    \STATE $p_i\leftarrow\mathbb{I}[u_i>0\land c_i=0]$, $r_i\leftarrow\mathbb{I}[\operatorname{inter\_repeat}_i=1\lor\operatorname{intra\_repeat}_i=1]$
    \STATE $z_i\leftarrow\mathbb{I}[p_i=1\lor r_i=1]$
    \STATE $(n_g,u_g,u)\leftarrow(n_g+z_i,u_g+z_i u_i,u+u_i)$
\ENDFOR
\STATE \textbf{\texttt{// Metric aggregation}}
\STATE $\operatorname{LGR}\leftarrow n_g/N$; $\operatorname{LGTS}\leftarrow u_g/\max(1,u)$
\RETURN $\{z_i\}_{i=1}^{N}$, $\operatorname{LGR}$, $\operatorname{LGTS}$
\end{algorithmic}
\end{algorithm}

Algorithm~\ref{alg:low_gain} provides the offline diagnostic counterpart of the online controller. The input records, variable domains, counters, and initial values are specified in lines 1--3. The risk-detection stage derives the post-tool failure flag $p_i$, repetition-risk flag $r_i$, and combined low-value execution flag $z_i$, while simultaneously accumulating flagged requests and tool calls (lines 4--9). The final stage aggregates LGR and LGTS and returns the request-level labels and diagnostic metrics (lines 10--12). Unlike Algorithm~\ref{alg:agentloop}, this procedure is post-hoc and characterizes workload-level execution risk rather than selecting the next online action.

\begin{figure*}[t]
  \centering
  \ecdfmethodlegend
  \par\vspace{0.5mm}
  \makebox[\textwidth][c]{%
    \ecdfpanel{0.31\textwidth}{0.10\textheight}{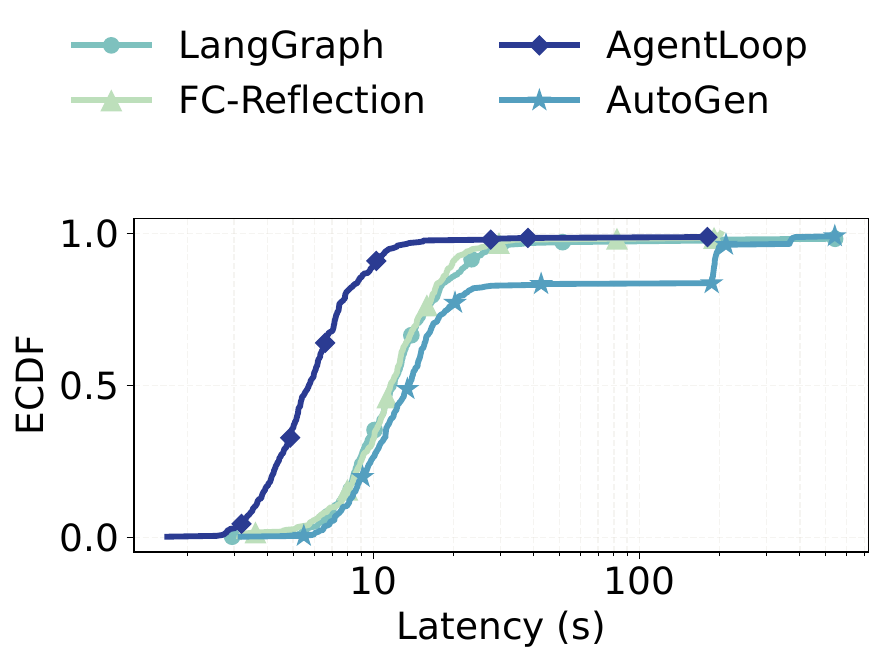}{(a) Llama: ToolBench.}%
    \hspace{0.02\textwidth}%
    \ecdfpanel{0.31\textwidth}{0.10\textheight}{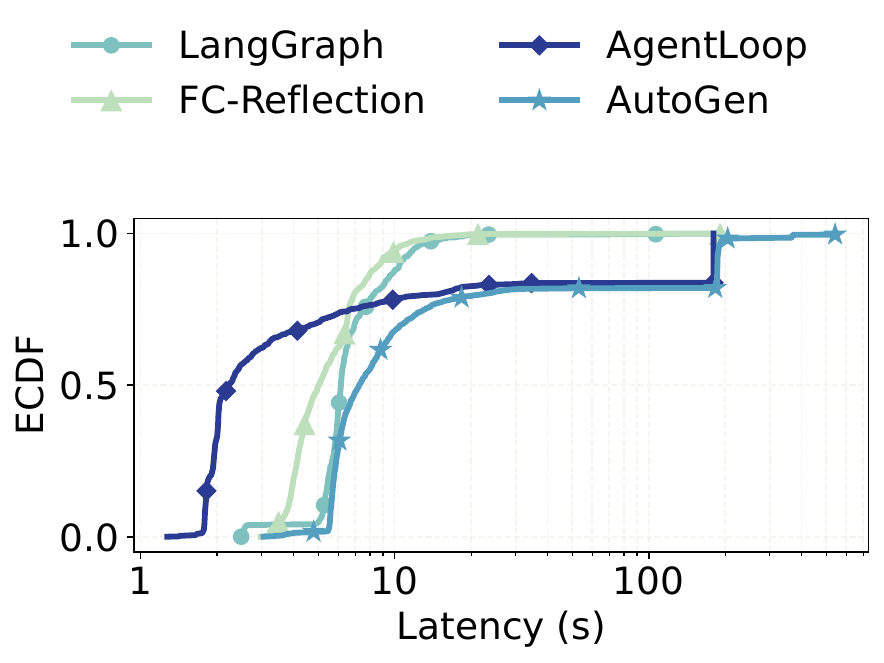}{(b) Llama: GAIA.}%
    \hspace{0.02\textwidth}%
    \wideecdfpanel{0.31\textwidth}{0.10\textheight}{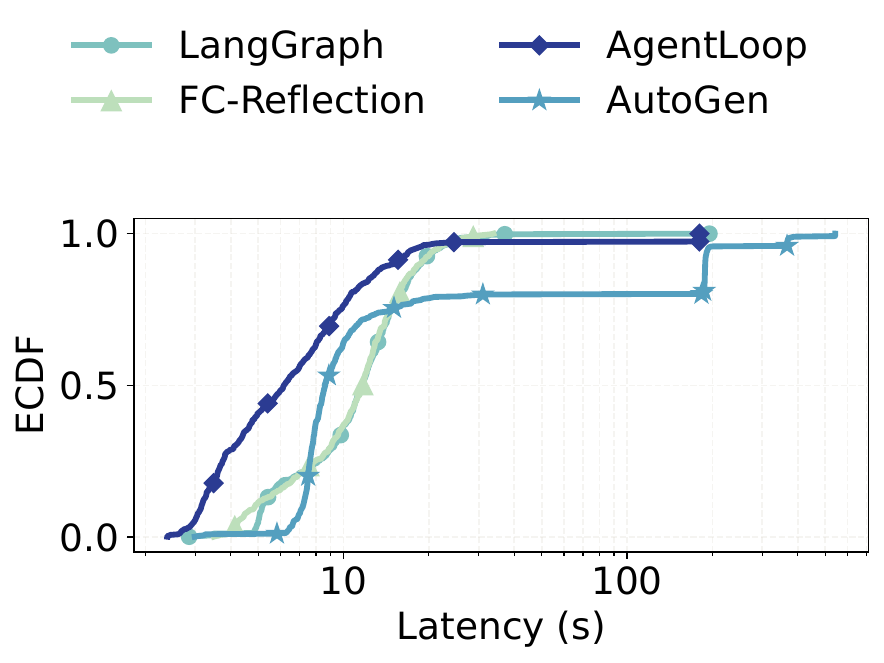}{(c) Llama: Open-Agent-Trace.}%
  }
  \par\vspace{0.5mm}
  \makebox[\textwidth][c]{%
    \qwenecdfpanel{0.31\textwidth}{0.10\textheight}{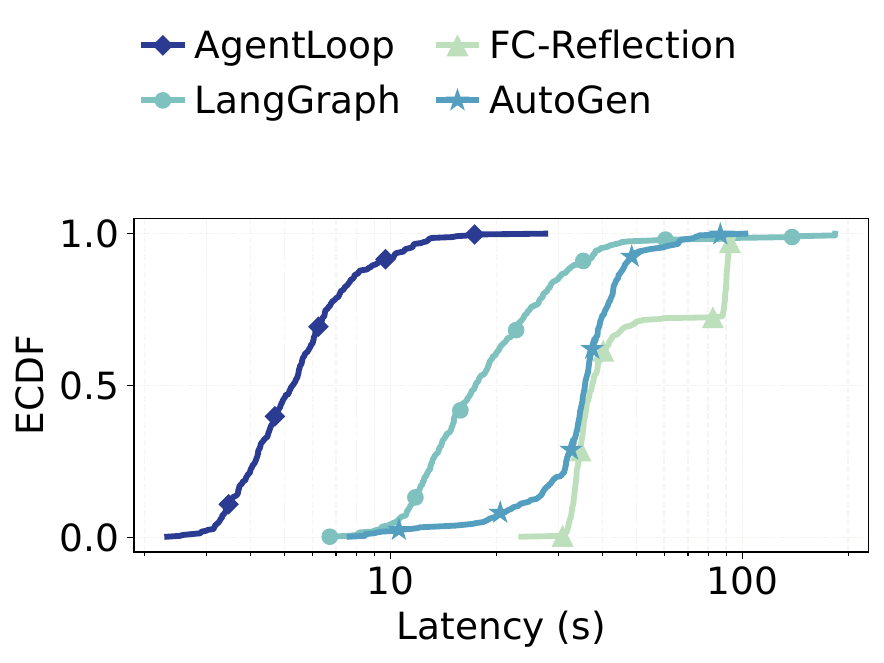}{(d) Qwen: ToolBench.}%
    \hspace{0.02\textwidth}%
    \qwenecdfpanel{0.31\textwidth}{0.10\textheight}{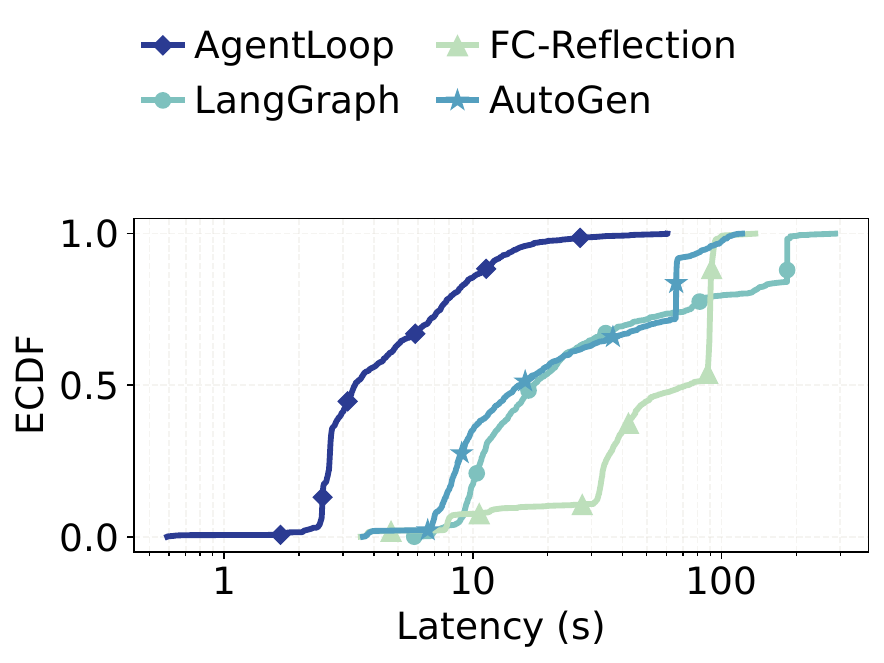}{(e) Qwen: GAIA.}%
    \hspace{0.02\textwidth}%
    \qwenecdfpanel{0.31\textwidth}{0.10\textheight}{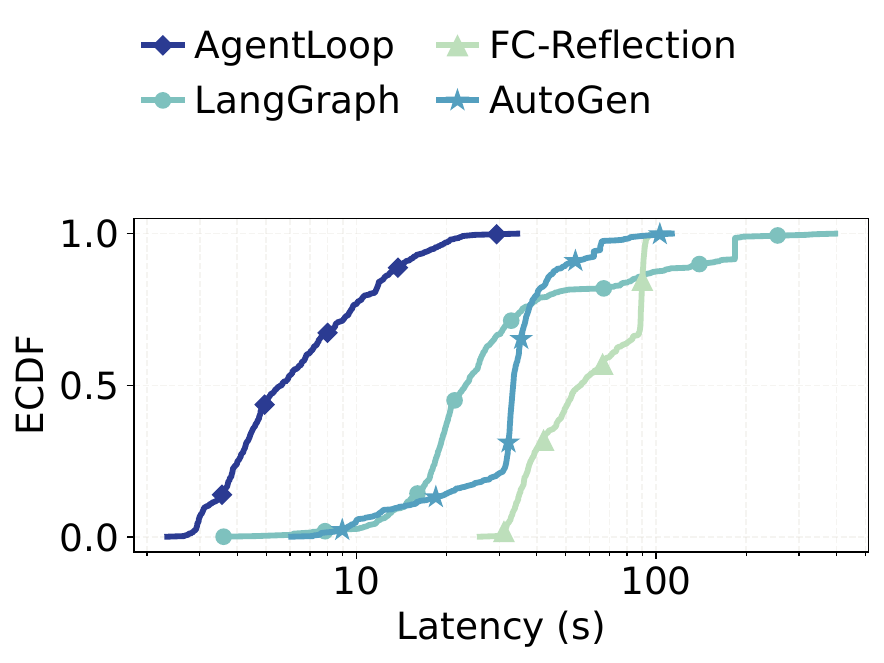}{(f) Qwen: Open-Agent-Trace.}%
  }
  \caption{End-to-end (E2E) latency ECDFs.}
  \label{fig:latency_ecdf}
\end{figure*}

\begin{figure*}[t]
  \centering
  \vspace{1mm}

  \breakdownlegend

  \vspace{0.5mm}

  \breakdownpanel{0.31\textwidth}{0.089\textheight}{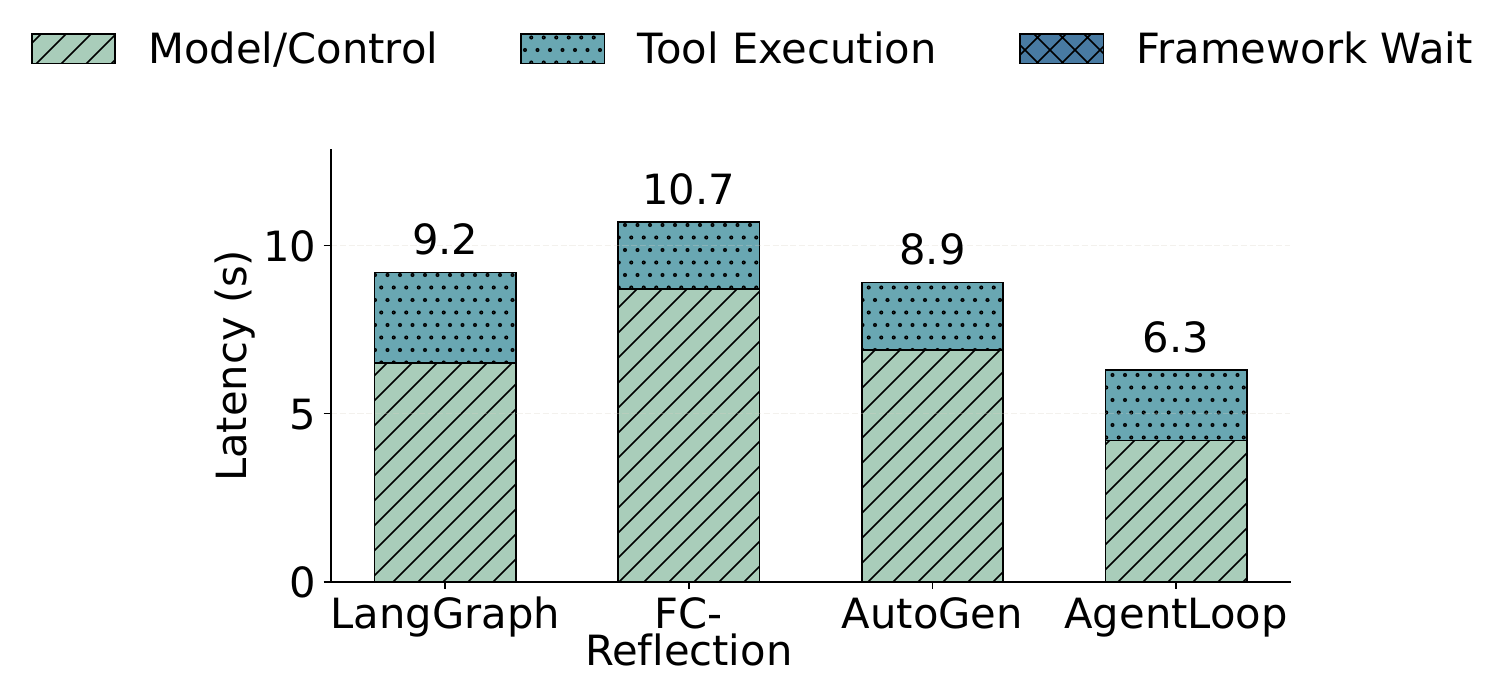}{(a) Llama: ToolBench.}
  \hfill
  \breakdownpanel{0.31\textwidth}{0.089\textheight}{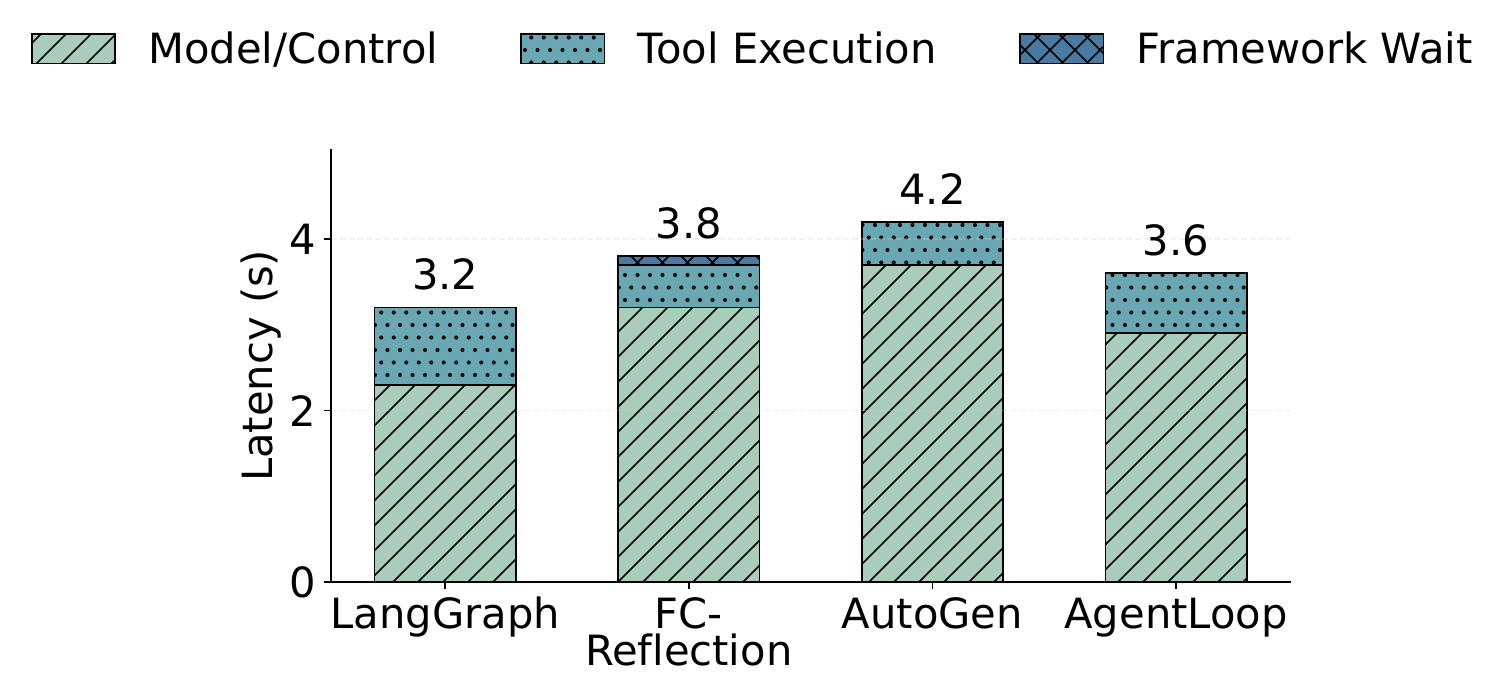}{(b) Llama: GAIA.}
  \hfill
  \breakdownpanel{0.31\textwidth}{0.089\textheight}{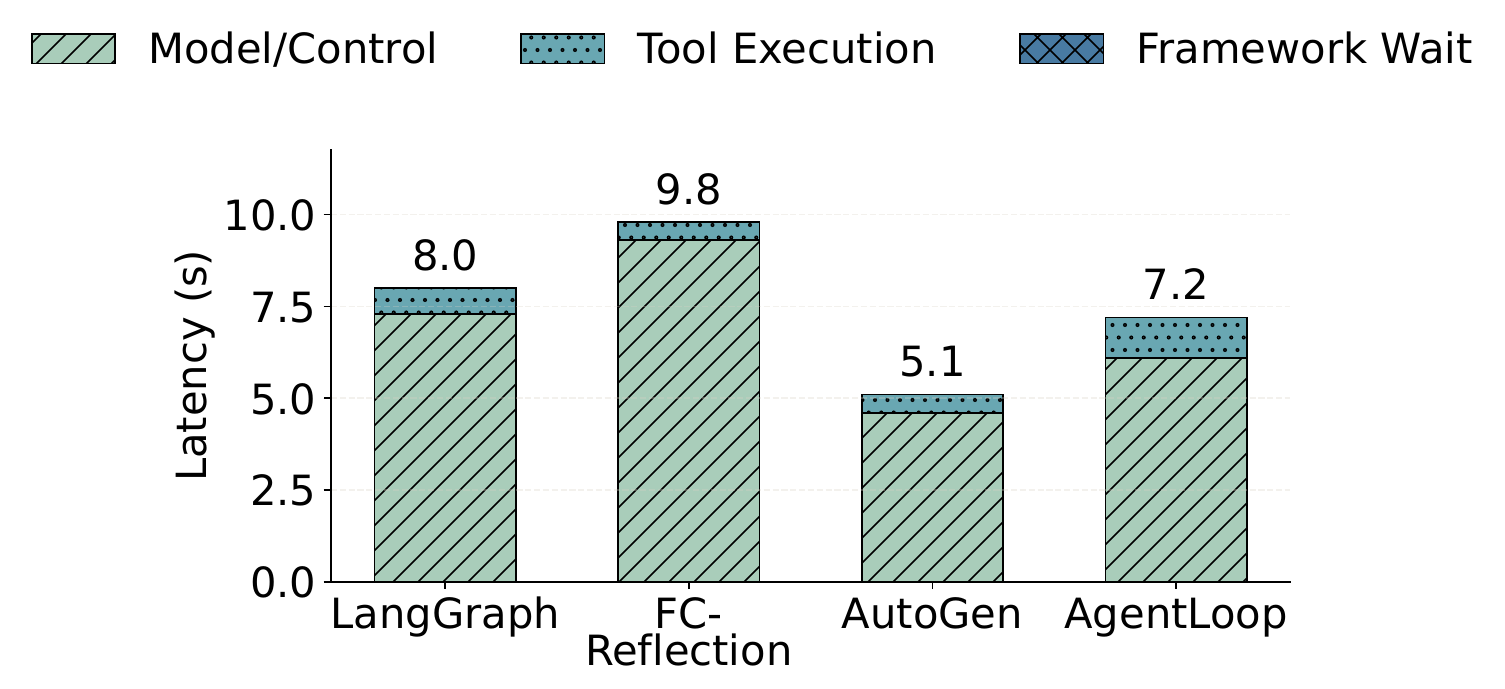}{(c) Llama: Open-Agent-Trace.}

  \vspace{0.5mm}

  \breakdownpanel{0.31\textwidth}{0.10\textheight}{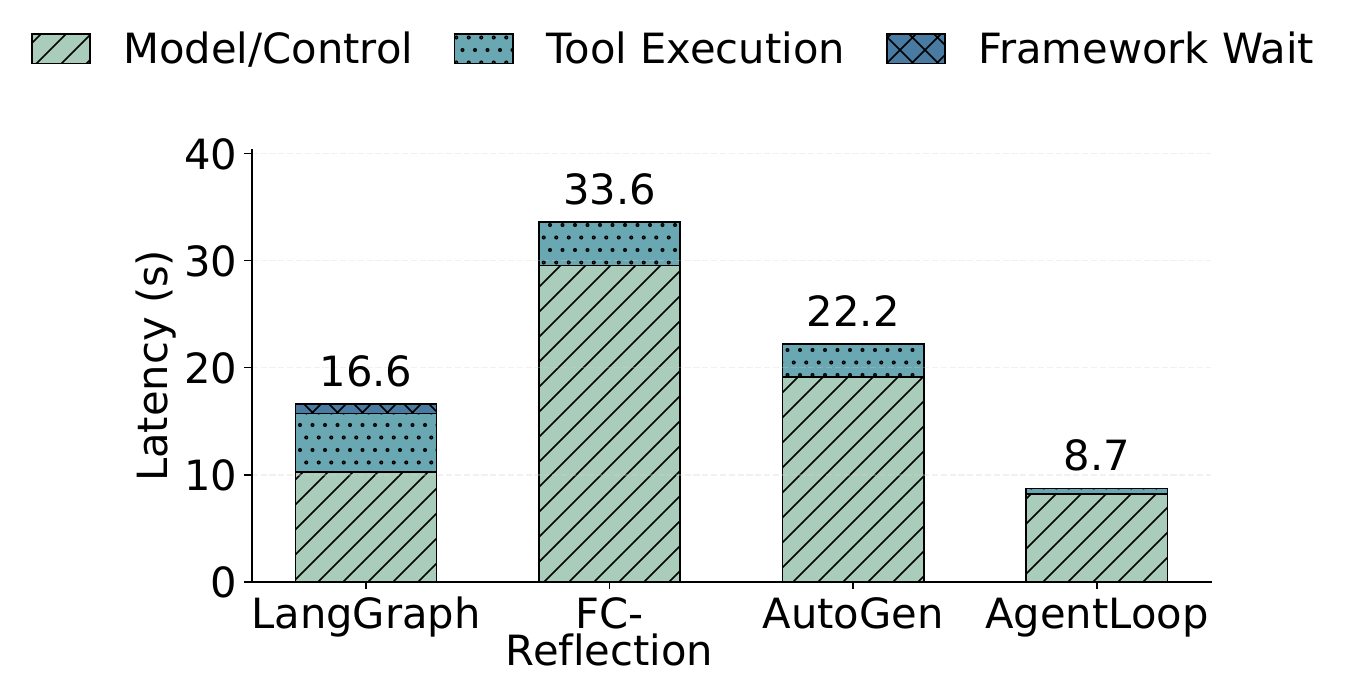}{(d) Qwen: ToolBench.}
  \hfill
  \breakdownpanel{0.31\textwidth}{0.10\textheight}{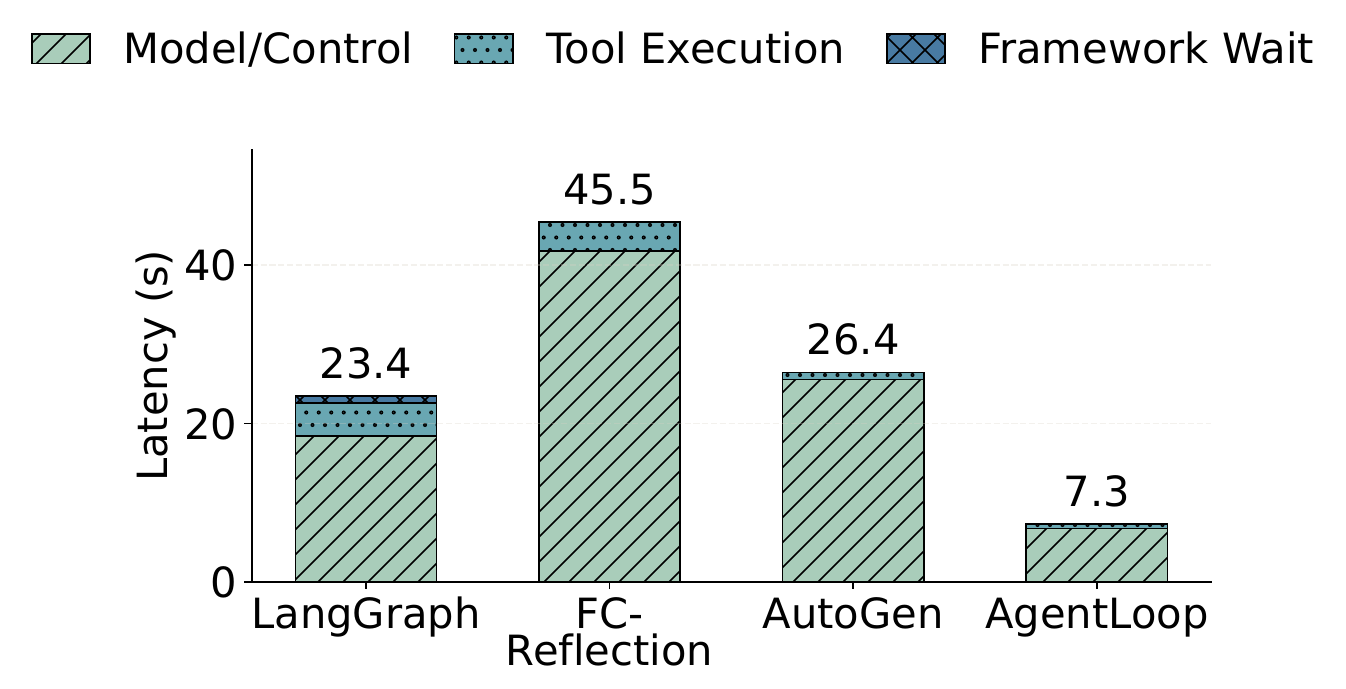}{(e) Qwen: GAIA.}
  \hfill
  \breakdownpanel{0.31\textwidth}{0.10\textheight}{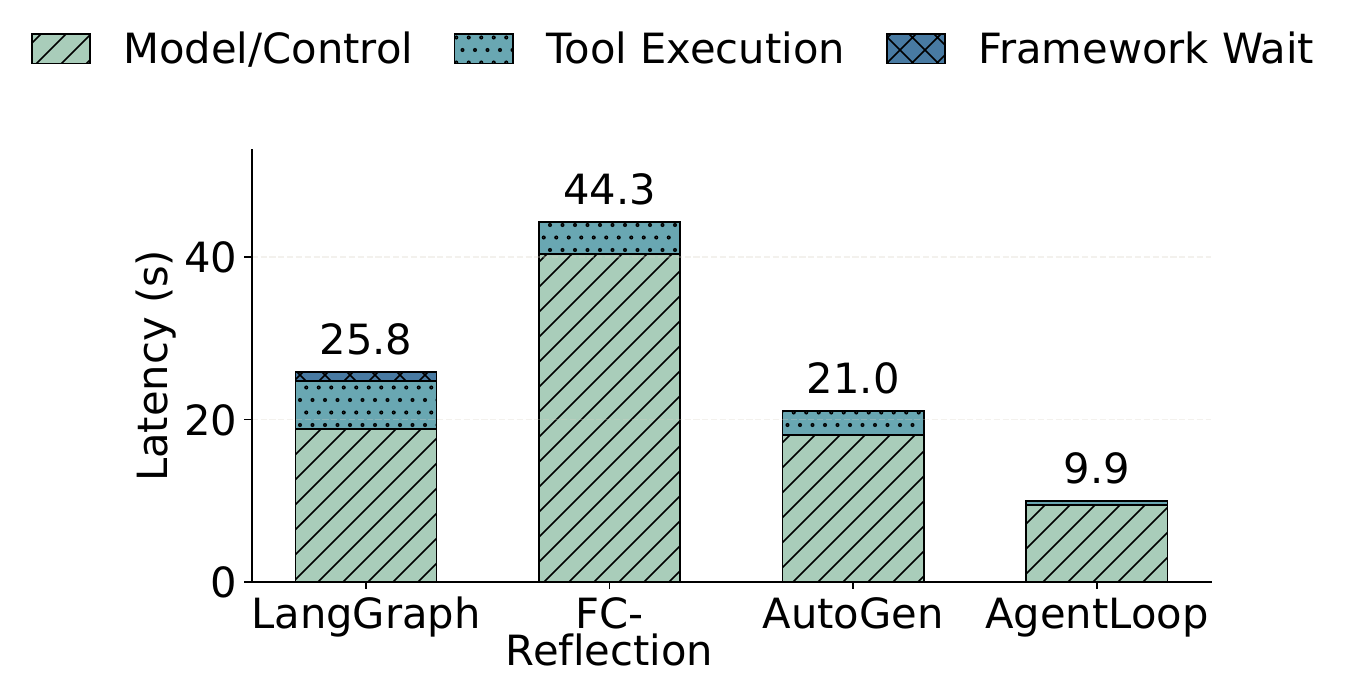}{(f) Qwen: Open-Agent-Trace.}
  \caption{End-to-end latency breakdown across model/control processing, tool execution, and framework waiting.}
  \label{fig:e2e_breakdown}
\end{figure*}

\subsection{Experimental Metrics and Complexity}

The execution graph closure rate (GCR) is an evaluation metric rather than an online control input:
\begin{equation}
\operatorname{GCR}=\frac{1}{N}\sum_{i=1}^{N}\mathbb{I}\!\left[\substack{ok_i=1\land\operatorname{NonEmptyOutput}_i=1\\
\land(\operatorname{NoTool}_i=1\lor\operatorname{NoPendingToolCall}_i=1)}\right].
\label{eq:gcr}
\end{equation}
Here, $ok_i$ indicates successful execution, $\operatorname{NonEmptyOutput}_i$ indicates that the final answer is nonempty, $\operatorname{NoTool}_i$ indicates that no tool was used, and $\operatorname{NoPendingToolCall}_i$ indicates that the final message has no pending tool call. GCR measures whether the execution graph returns to a final answer state. It is not an accuracy metric.

At the event level, the stable tool-set stop rate (STSSR) is
\begin{equation}
\operatorname{STSSR}=\frac{\sum_{j=1}^{J}\mathbb{I}[action_j=\text{stop\_replanning}]}{\max(1,J)}.
\label{eq:stssr}
\end{equation}
Here, $J$ is the number of stability-gate checks and $action_j$ is the action emitted at check $j$. This event-level metric should be distinguished from a request-level posterior stable-stop rate that may be computed by a separate analysis script.

\textbf{Complexity Analysis.} Let $T_{\mathrm{main}}$ be the actual number of main-loop tool rounds, with $T_{\mathrm{main}}\le T_{\max}$, and let $T_{\mathrm{ref}}$ be the number of extra reflection tool rounds. Let $R_{\max}$ be the maximum reflection rounds, $P_{\max}$ the maximum calls in one tool round, $L_{\mathrm{arg}}$ and $L_{\mathrm{res}}$ the average argument and result lengths, and $I_{\max}$ the maximum answer-integration or recheck count. Let $C_{\mathrm{plan}}$ be one planning-call cost, $C_{\mathrm{tool\_wall}}$ one round of tool wall-clock cost, $C_{\mathrm{verify}}$ one verifier-call cost, $C_{\mathrm{integrate}}$ one answer-integration cost, and $C_{\mathrm{control}}$ the local cost of control, signature construction, and gating. Let $H$ be the context and history space, and let $U$ be the number of different call keys in the tool cache. Then
\begin{equation}
\resizebox{0.98\columnwidth}{!}{$
\begin{aligned}
\operatorname{Time}(\mathrm{AgentLoop})={}&O\!\bigl((T_{\mathrm{main}}+T_{\mathrm{ref}})
 [C_{\mathrm{plan}}+C_{\mathrm{tool\_wall}}\\
&\qquad+P_{\max}(L_{\mathrm{arg}}+L_{\mathrm{res}})]
 +R_{\max}[C_{\mathrm{verify}}+C_{\mathrm{control}}]\\
&\qquad+I_{\max}[C_{\mathrm{integrate}}+C_{\mathrm{verify}}]\bigr),\\
\operatorname{Space}(\mathrm{AgentLoop})={}&O\!\left(\begin{gathered}
H+U(L_{\mathrm{arg}}+L_{\mathrm{res}})\\
+P_{\max}(L_{\mathrm{arg}}+L_{\mathrm{res}})
\end{gathered}\right).
\end{aligned}
 $}
\label{eq:complexity}
\end{equation}
The tool calls in one round may run concurrently, so $C_{\mathrm{tool\_wall}}$ is close to the wall time of the slowest tool. Total resource consumption still accumulates the cost of all tool calls. LLM inference and external service execution are black-box service costs and should not be reduced to ordinary string-length $O(n)$ terms. AgentLoop's added local work mainly comes from argument and result normalization, sorting, signature construction, and similarity computation. If the tool-state cache is disabled, the cache term containing $U$ can be removed, but the history context $H$ remains.
\section{Performance Evaluation}

\textbf{Datasets and Tasks.} The evaluation covers thousands of requests across three datasets. ToolBench emphasizes multi-API selection and composition, GAIA emphasizes grounded QA, and Open-Agent-Trace emphasizes realistic multi-step workflows.

\textbf{ToolBench}\cite{qin2024toolllm} contains 356 tool-intensive requests focused on API selection and multi-service composition .

\textbf{GAIA}\cite{shamikbose89_gaia_traces} denotes the evaluated subset of GAIA-traces. The original collection contains 1,204 grounded question-answering trace requests, from which we select 1,000 requests with complete task and trace information for evaluation.

\textbf{Open-Agent-Trace}\cite{simon2025openagenttraces} contains 499 requests representing realistic multi-step agent workflows .

All methods use the same datasets, concurrency setting, timeout boundary, and cache-disabled configuration. We evaluate Llama-3-8B-Instruct and Qwen3.6-35B-A3B under the same agent settings.

\textbf{Baselines and Metrics.} We compare AgentLoop with AutoGen\cite{wu2023autogen}, FC-Reflection\cite{schick2023toolformer,shinn2023reflexion}, and LangGraph\cite{langchain2026langgraph}.

Evaluation metrics include \textit{average E2E latency}, \textit{FTR}, \textit{average iteration depth}, \textit{tool calls per request}, \textit{tokens per request}, \textit{ECDFs}, \textit{average latency}, \textit{effective-token redundancy}, and \textit{answer accuracy}. Execution success measures whether the process completes, while answer accuracy measures whether the final response is supported by evidence, so the two are reported separately rather than merged into a single score.

\textbf{Experimental Configurations.} Experiments were conducted on a unified GPU node with four NVIDIA Tesla V100 PCIe GPUs, each with 32 GB of memory. All methods used the same hardware platform and model service. Inference used a unified local model service. The main experiments used Llama-3-8B-Instruct, and the cross-model setting used Qwen3.6-35B-A3B with GPTQ Int4 and vLLM FP16. The request rate was 1 request/s, the timeout was 180s, and cache was disabled. For each dataset and method, we used the identical request order and runtime parameters. The experimental results analysis are as below: 

\deferfloatstonextpage
\begin{figure}[t]
  \centering
  \ftrmethodlegend

  \vspace{-0.5mm}

  \ftrpanel{0.47\linewidth}{64pt}{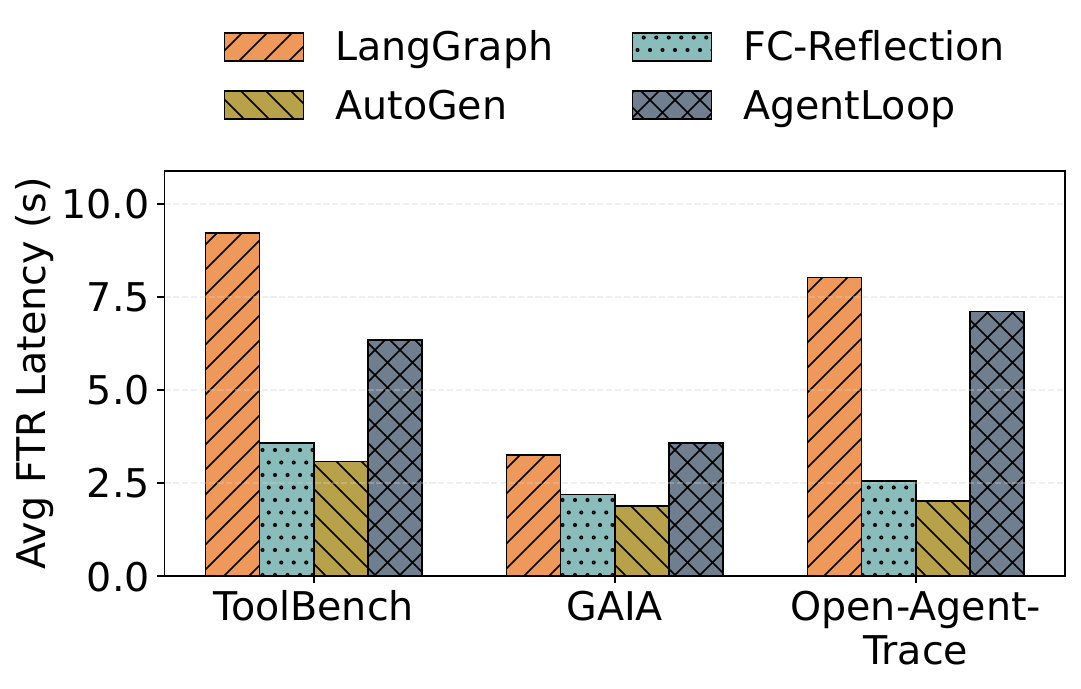}{(a) Average FTR comparison.}
  \hfill
  \ftrpanel{0.47\linewidth}{64pt}{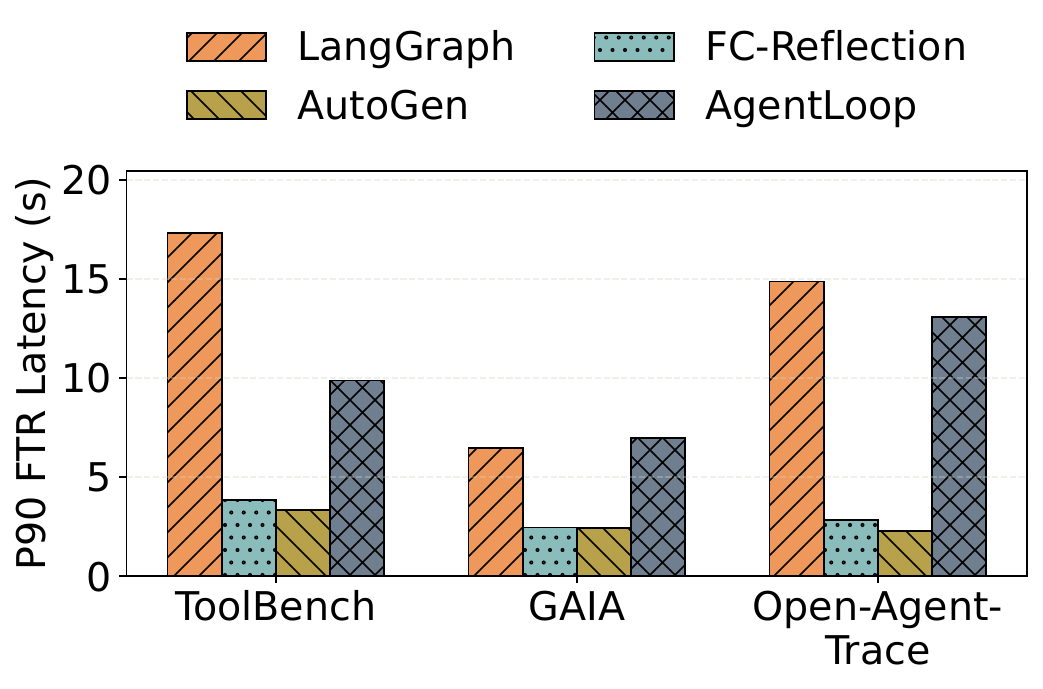}{(b) P90 FTR comparison.}
  \caption{Average and P90 first-token response latency for Llama.}
  \label{fig:ftr_llama}
\end{figure}

\begin{figure}[t]
  \centering
  \ftrmethodlegend

  \vspace{-0.5mm}

  \ftrpanelwide{0.47\linewidth}{64pt}{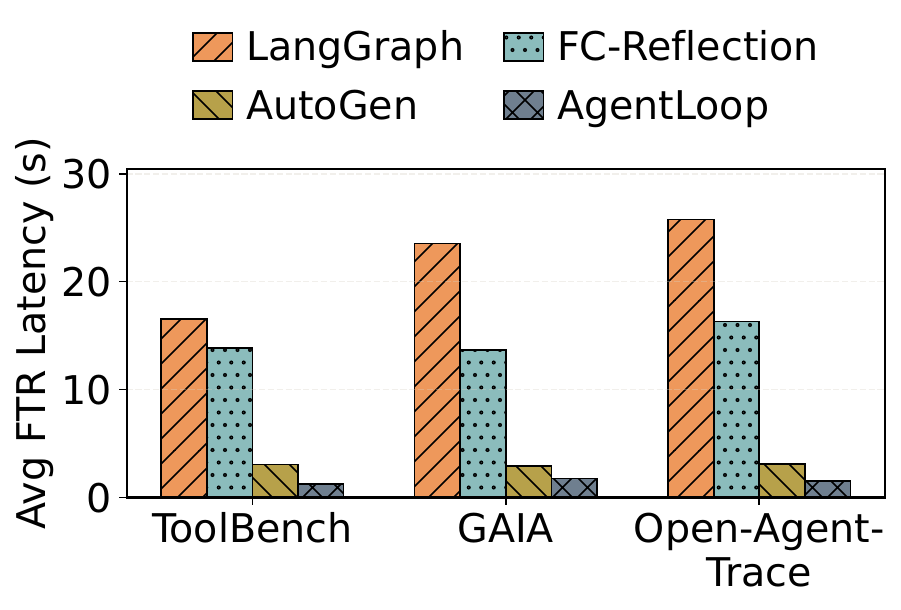}{(a) Average FTR comparison.}
  \hfill
  \ftrpanelwideb{0.47\linewidth}{64pt}{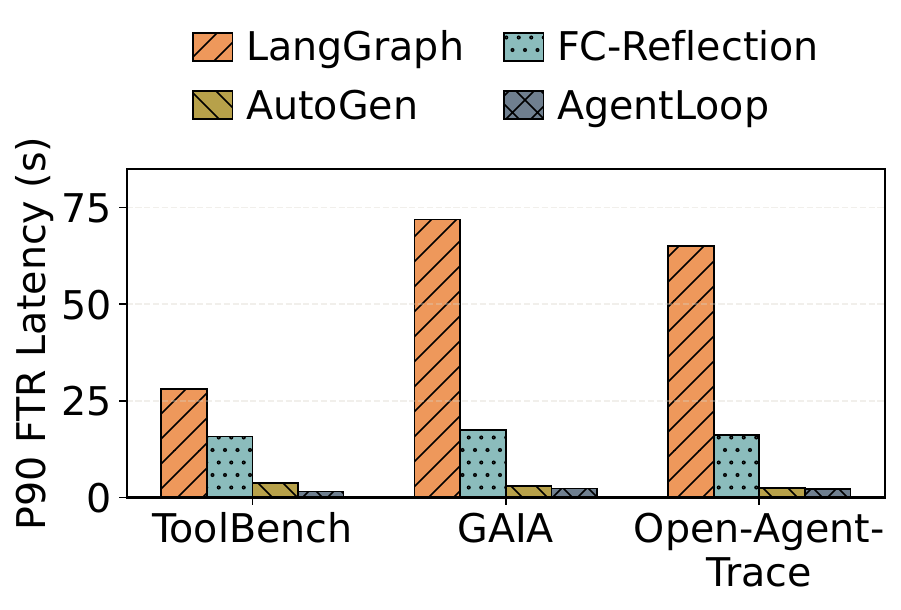}{(b) P90 FTR comparison.}
  \caption{Average and P90 first-token response latency for Qwen.}
  \label{fig:ftr_qwen}
\end{figure}

\paragraph{\textbf{Latency Comparison}}

To compare latency distribution, Fig.~\ref{fig:latency_ecdf} demonstrates that AgentLoop generally shifts requests toward lower E2E latency, especially on ToolBench and Open-Agent-Trace. In the Llama setting, GAIA retains a heavier long tail, whereas the Qwen setting shows a broader advantage across all three datasets. AgentLoop reaches 8.787s average latency on Llama ToolBench and 15.144s P90 latency on Llama Open-Agent-Trace; on Qwen, its ToolBench average is about 5.85s and the largest average reduction appears on GAIA. On ToolBench, the average latency is reduced by more than 70\% relative to the stronger baselines. These gains result from slot-closure and stability checks that suppress repeated reasoning-action rounds, rather than from faster external tools. The remaining GAIA tail also shows that completion control cannot eliminate requests whose evidence is intrinsically difficult to obtain.

The same control effect appears in the breakdown and FTR results. Fig.~\ref{fig:e2e_breakdown} shows that the clearest reduction is in model/control processing, confirming that AgentLoop avoids repeated planning while leaving tool execution unchanged. The advantage is therefore a coordination gain: the controller reduces unnecessary rounds before they become additional model and framework waiting time. Figs.~\ref{fig:ftr_llama} and \ref{fig:ftr_qwen} further show earlier first-token responses on most datasets, with the clearest improvement against LangGraph; Qwen achieves the best FTR on all datasets. Thus, the latency benefit is shared across both backbones and includes both common-case and tail-response improvements.
In the Llama results, the gains are strongest on workloads with multi-step service composition, where repeated planning contributes a larger share of the total delay. The weaker GAIA result is consistent with requests that require more evidence acquisition before closure.

\paragraph{\textbf{Token and Execution Cost}}

\begin{figure*}[t]
  \centering
  \tokenmethodlegend

  \vspace{0.5mm}

  \tokenresultpanel{0.24\textwidth}{0.999\linewidth}{0.10\textheight}{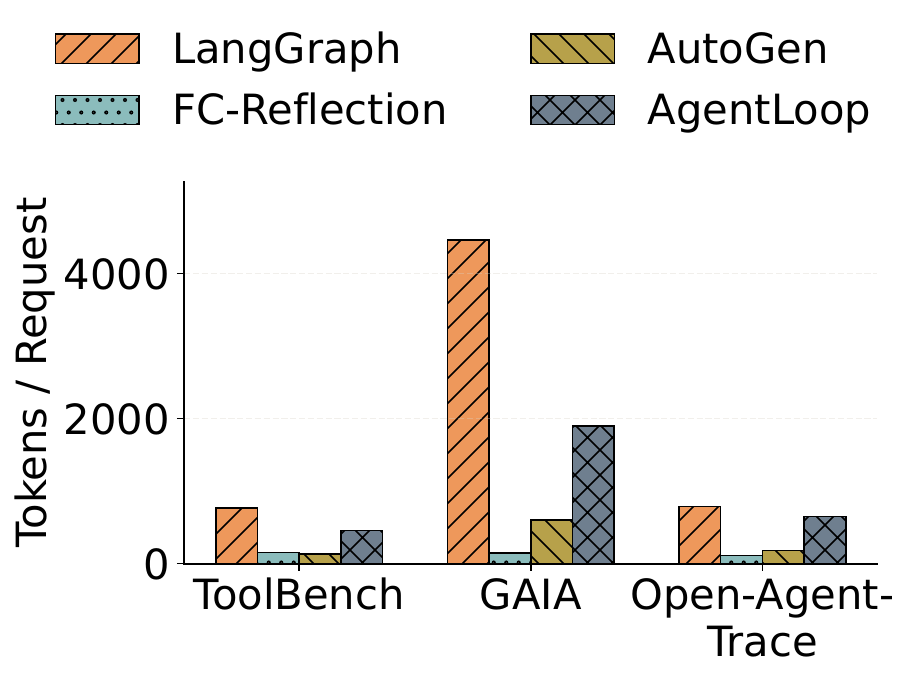}{(a) Llama: input tokens.}
  \hfill
  \tokenresultpanel{0.24\textwidth}{0.999\linewidth}{0.10\textheight}{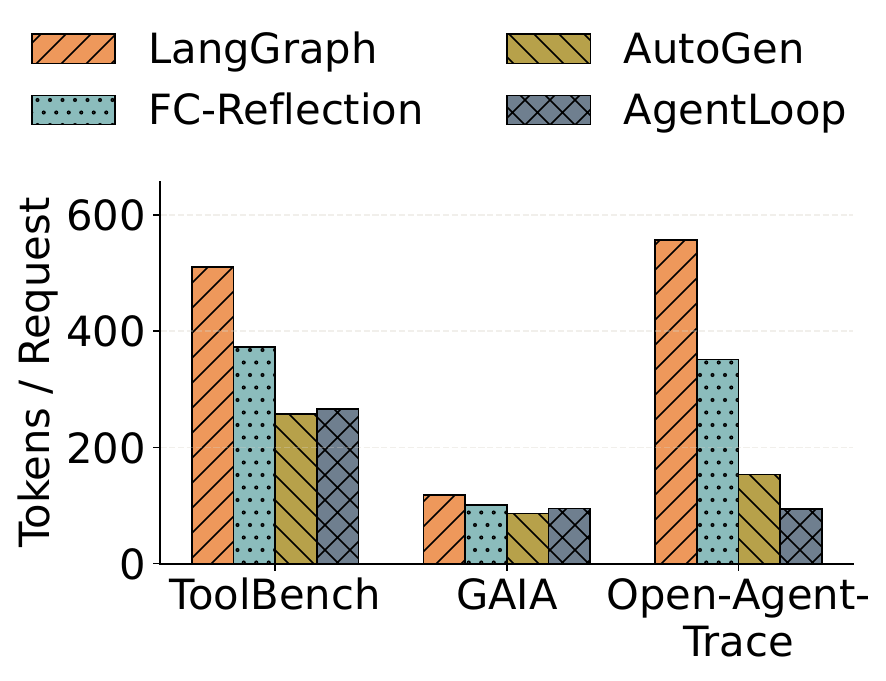}{(b) Llama: output tokens.}
  \hfill
  \smallqwentokenresultpanel{0.24\textwidth}{0.999\linewidth}{0.10\textheight}{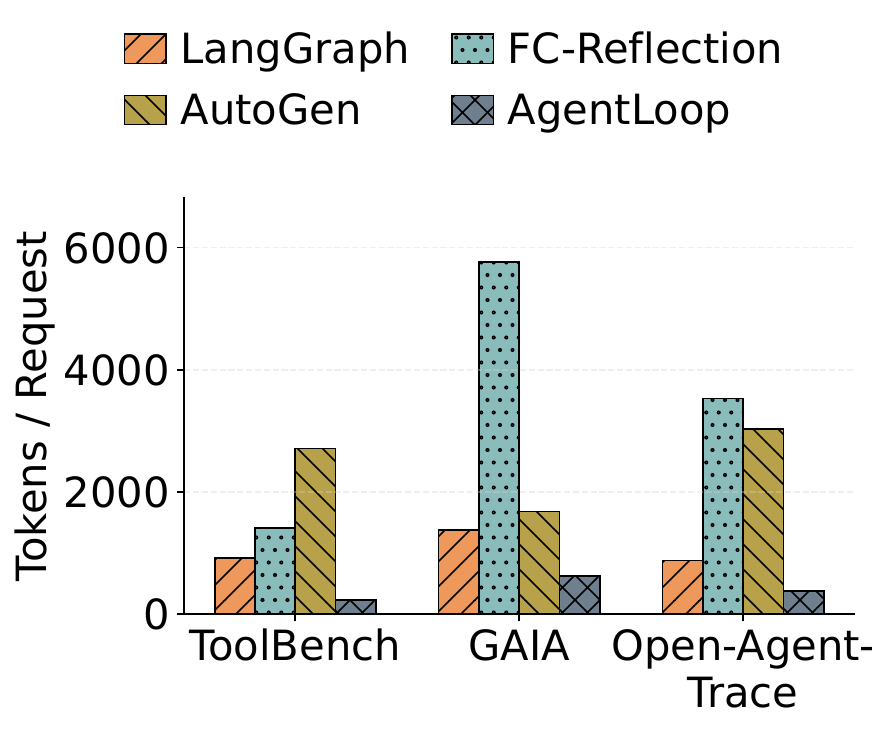}{(c) Qwen: input tokens.}
  \hfill
  \smallqwentokenresultpanel{0.24\textwidth}{0.999\linewidth}{0.10\textheight}{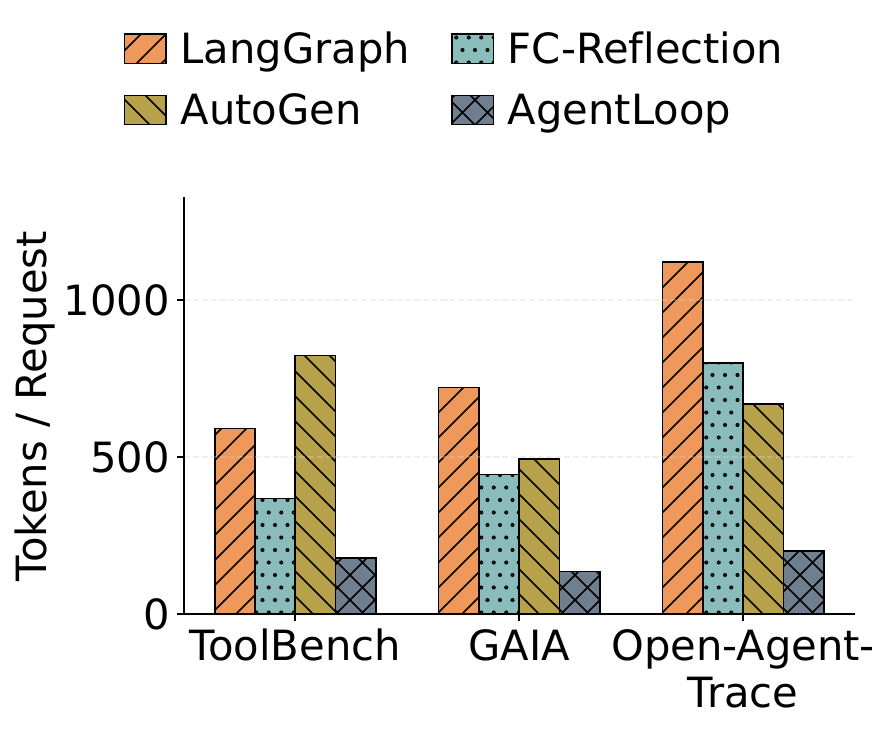}{(d) Qwen: output tokens.}
  \caption{Input and output token cost across Llama and Qwen.}
  \label{fig:token_cost}
\end{figure*}

To compare token and execution cost, Fig.~\ref{fig:token_cost} demonstrates a consistent reduction in input and output tokens, comp baselines. AgentLoop achieves this mainly through shorter reasoning-action loops and a compact runtime state instead of repeatedly reassembling the full trace. The benefit is larger on longer traces because, once the required slots are covered, carrying the complete interaction history contributes overhead rather than useful state. The reduction is substantial against LangGraph on Llama, particularly on ToolBench and GAIA, and carries over to Qwen, where AgentLoop has the lowest input and output token costs on all three datasets.
The two token views also separate prompt-side savings from response-side savings. Their consistent direction suggests that the reduction is not caused by truncating only generated answers, but by avoiding repeated context construction and unnecessary loop expansions. This is important for long-horizon service workflows, where input-token accumulation can dominate the total cost.
It also shows that the control state is not merely a stopping signal, but a compact representation that reduces the amount of repeated evidence passed into later model calls.

\begin{figure}[t]
  \centering
  \includegraphics[width=0.7\linewidth]{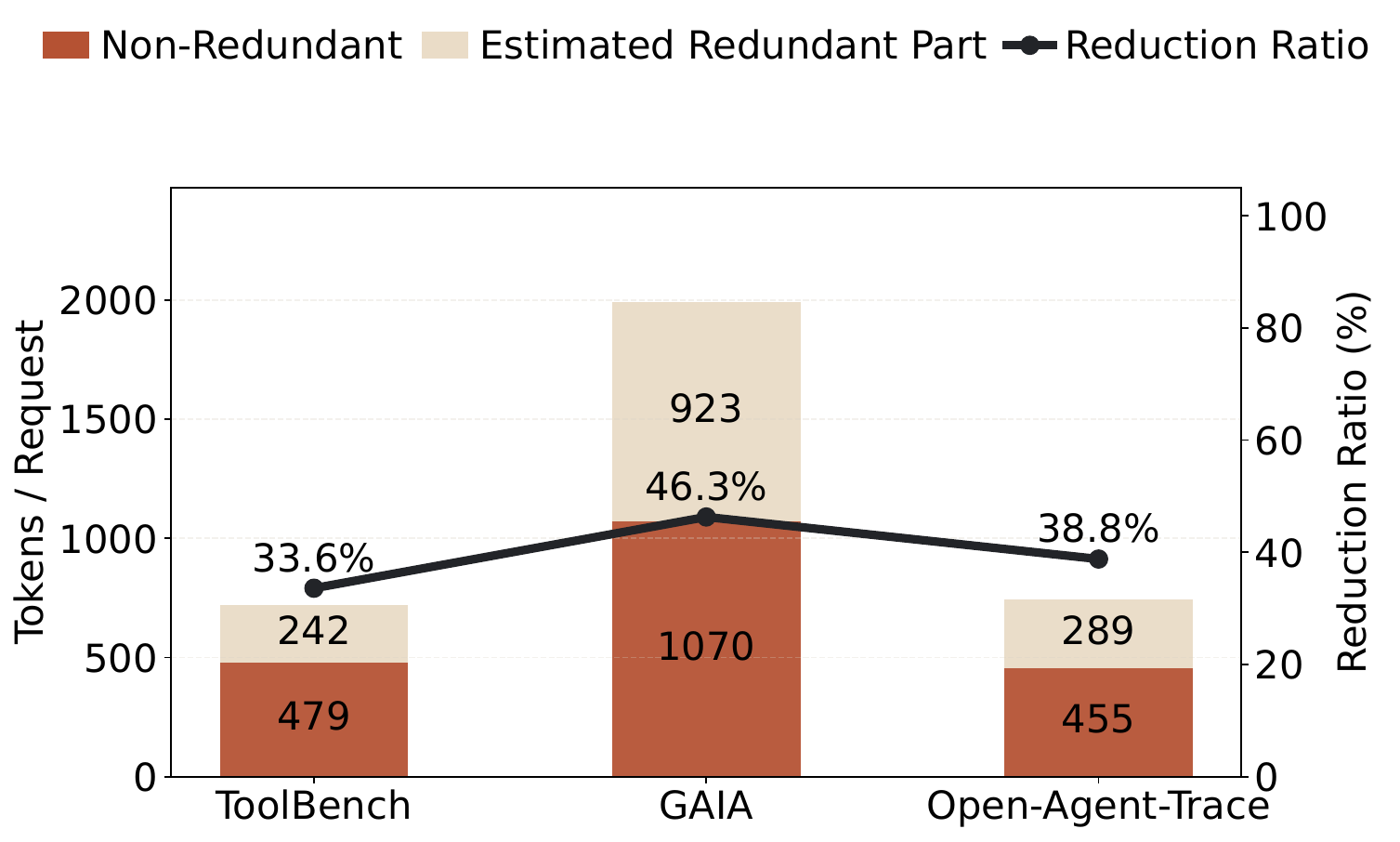}
  \caption{Control-side effective-token redundancy for Llama.}
  \label{fig:llama_redundancy}
\end{figure}

\begin{figure}[t]
  \centering
  \resourcedepthlegend

  \vspace{1mm}

  \resourcepanel{0.49\columnwidth}{0.132\textheight}{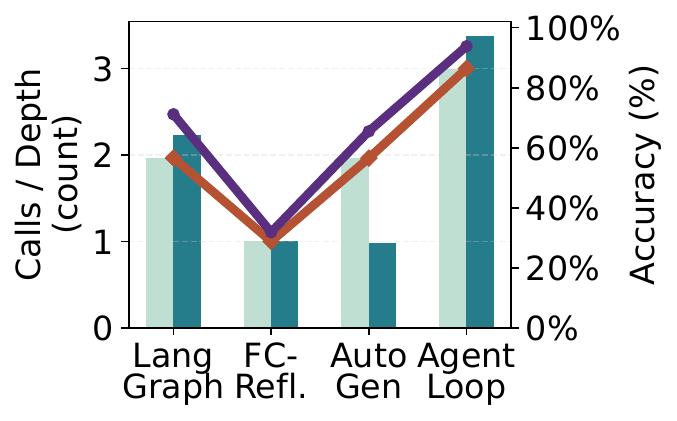}{(a) Llama: ToolBench.}
  \hfill
  \qwenresourcepanel{0.49\columnwidth}{0.132\textheight}{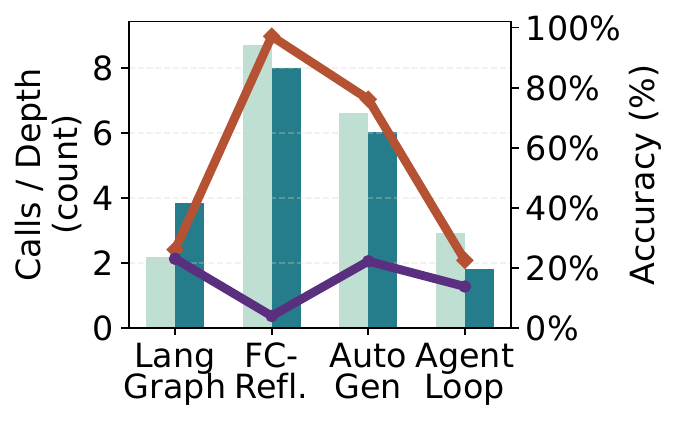}{(b) Qwen: ToolBench.}

  \vspace{-3mm}

  \resourcepanel{0.49\columnwidth}{0.132\textheight}{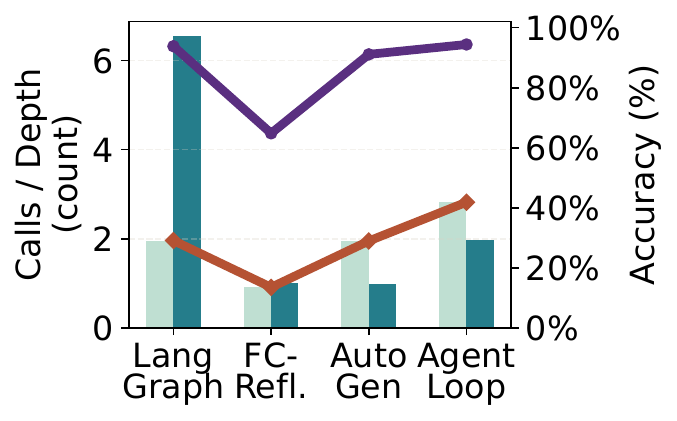}{(c) Llama: GAIA.}
  \hfill
  \qwenresourcepanel{0.49\columnwidth}{0.132\textheight}{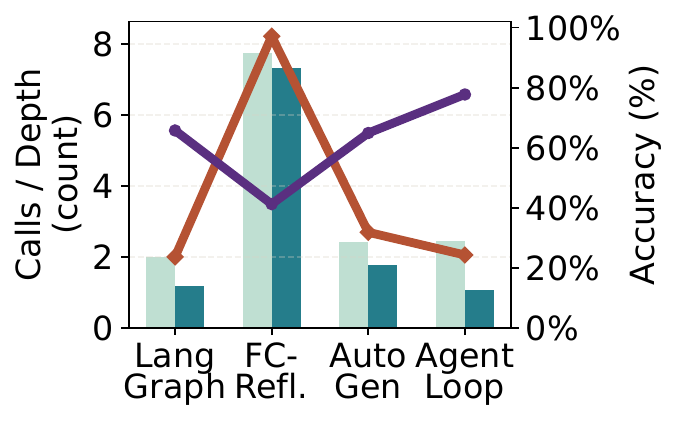}{(d) Qwen: GAIA.}

  \vspace{-3mm}

  \resourcepanel{0.49\columnwidth}{0.132\textheight}{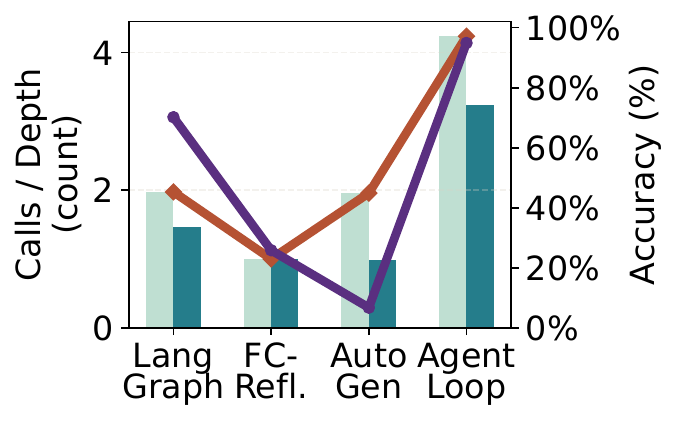}{(e) Llama: Open-Agent-Trace.}
  \hfill
  \qwenresourcepanel{0.49\columnwidth}{0.132\textheight}{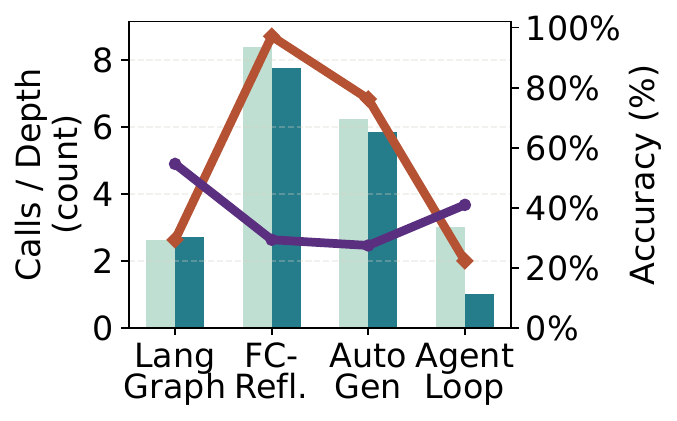}{(f) Qwen: Open-Agent-Trace.}
  \caption{Execution depth, LLM calls, tool calls, and answer accuracy across Llama and Qwen. Each row pairs the two backbones on the same dataset, with Llama in the left column and Qwen in the right column. Horizontal-axis method labels are wrapped over two lines, where FC-Refl.\ abbreviates FC-Reflection.}
  \label{fig:resource_depth}
\end{figure}

Fig.~\ref{fig:resource_depth} combines execution depth, LLM calls, tool calls, and answer accuracy. A shorter bar group with a higher accuracy point is the desired efficiency-quality pattern. AgentLoop keeps depth moderate while preserving high accuracy in the Llama panels, staying near a depth of three on ToolBench and retaining the higher accuracy curve with a shorter chain on Open-Agent-Trace. In the Qwen panels, it is not always the most accurate method, but consistently shortens execution by making continuation conditional on unresolved slots and positive marginal utility; on GAIA, it can lead in accuracy while using far fewer steps. The result indicates selective stopping rather than depth reduction alone.
The comparison also clarifies that execution depth should not be interpreted as a quality objective by itself. AgentLoop continues when the slot state or evidence remains unresolved, but avoids spending additional depth after the state has become sufficiently stable. This explains why its efficiency gain can coexist with competitive accuracy.
Although the absolute depth and accuracy vary between Llama and Qwen, the same conditional-continuation principle is visible in both settings.
This behavior is especially important for service-computing workloads, where excessive depth increases not only model cost but also external invocation latency and scheduling pressure. By linking continuation to unresolved slots, AgentLoop reduces these costs while preserving the execution capacity needed for genuinely multi-step requests.

\paragraph{\textbf{Redundancy Suppression and Slot Closure}}

\begin{table}[!t]
  \caption{Controller-side control metrics of AgentLoop. }
  \label{tab:internal_control_metrics}
  \centering
  \footnotesize
  \setlength{\tabcolsep}{4.5pt}
  \begin{tabular}{@{}lccc@{}}
  \toprule
  Dataset & Slot Closure & Stable Hit & Stable Stop \\
  \midrule
  ToolBench & 98.6\% & 97.3\% & 96.9\% \\
  GAIA & 81.6\% & 78.4\% & 77.9\% \\
  Open-Agent-Trace & 95.6\% & 91.4\% & 91.0\% \\
  \bottomrule
  \end{tabular}
  
     {\footnotesize \textit{Note:} The table shows whether the controller marks required slots as closed, detects stable evidence or action states, and converts stability into stopping or answer synthesis. All three quantities are recorded by the controller itself and describe its internal decision behavior, not the correctness of the answer.}
     
\end{table}

To examine the effects of runtime controller, Table~\ref{tab:internal_control_metrics} reports Slot Closure, Stable Hit, and Stable Stop as controller-side control metrics. Let $N$ be the number of requests, let $\mathcal{V}_i$ denote the slot-verification events recorded for request $i$, and let $\mathcal{G}_i$ denote the stability-gate checks recorded for request $i$. Slot Closure is the fraction of requests whose last verification event reports a closed runtime state, that is, $C_t=1$ in Equation \eqref{eq:closure}:
\begin{equation}
\operatorname{SlotClosure}=\frac{1}{N}\sum_{i=1}^{N}\mathbb{I}\!\left[\mathcal{V}_i\neq\emptyset\land a^{(i)}_{\mathrm{last}}=1\land n^{(i)}_{\mathrm{last}}=0\right].
\label{eq:slot_closure_rate}
\end{equation}
Stable Hit is the fraction of requests in which the stability gate was reached at least once, so it measures gate coverage rather than the outcome of the comparison:
\begin{equation}
\operatorname{StableHit}=\frac{1}{N}\sum_{i=1}^{N}\mathbb{I}\!\left[\mathcal{G}_i\neq\emptyset\right].
\label{eq:stable_hit_rate}
\end{equation}
Stable Stop is the fraction of requests in which stability was converted into an early stop of replanning:
\begin{equation}
\resizebox{0.98\columnwidth}{!}{$
\operatorname{StableStop}=\frac{1}{N}\sum_{i=1}^{N}\mathbb{I}\!\left[\exists\, g\in\mathcal{G}_i:\ action_g=\text{stop\_replanning}\right].
$}
\label{eq:stable_stop_rate}
\end{equation}
All three metrics use the request count $N$ as denominator. Since the gate exits replanning once it fires, $\operatorname{StableStop}\le\operatorname{StableHit}$, and their gap indicates reached gates that did not trigger early stopping. Stable Stop is request-level, whereas STSSR in Equation \eqref{eq:stssr} is event-level. Slot Closure is 98.6\% on ToolBench, 81.6\% on GAIA, and 95.6\% on Open-Agent-Trace; Stable Hit and Stable Stop are also close, indicating that reached gates usually produce early stops. These values describe controller behavior rather than final-answer correctness, which is evaluated separately. In a stratified hand-labeled sample of 96 requests aligned with verifier decisions, precision, recall, and F1 were 0.840, 0.913, and 0.875, respectively. Because the sample over-represents difficult cases, these values characterize verifier behavior on hard requests rather than the average request.
The lower closure rate on GAIA is consistent with its longer-tail behavior: more requests finish with unresolved evidence even though the controller still detects stable states in a substantial fraction of cases. The close Stable Hit and Stable Stop values on all datasets indicate that the stability signal is usually converted into an action instead of remaining only an internal observation.
This separation also prevents the controller-side metrics from being mistaken for accuracy measurements: they explain why the loop stops, while the final answer quality is evaluated by the accuracy results above.
This distinction is important because an early stop can be desirable only when it is supported by slot closure or stable evidence, not when it simply reflects an exhausted budget or an arbitrary workflow boundary.

\begin{table}[!t]
  \caption{Measured slot-verification overhead of AgentLoop. }
  \label{tab:slot_overhead}
  \centering
  \footnotesize
  \setlength{\tabcolsep}{4.5pt}
  \begin{tabular}{@{}lcc@{}}
  \toprule
  Dataset & Slot Verification & Avg.\ Verification \\
          & Time Share & Latency (ms) \\
  \midrule
  ToolBench & 3.61\% & 317.4 \\
  GAIA & 0.40\% & 129.9 \\
  Open-Agent-Trace & 1.72\% & 206.6 \\
  \bottomrule
  \end{tabular}
  
   {\footnotesize \textit{Note:} Time share is the total slot-verification time divided by the total end-to-end latency of the same requests, and latency is the per-request verification time, that is, the verification time accumulated within a request and averaged over all requests.}
\end{table}

Table~\ref{tab:slot_overhead} instantiates the $C_{\mathrm{verify}}$ term of Equation~\eqref{eq:complexity}. Slot verification accounts for at most 3.61\% of E2E latency, with per-request verification time between 129.9 ms and 317.4 ms. ToolBench has the highest share (3.61\%) because its shorter chains provide less amortization, while GAIA has the lowest share (0.40\%) because its longer requests amortize verification overhead. Thus, the control layer remains a second-order cost relative to model inference and tool execution, although the fixed cost is more visible for very short tool chains. The trade-off is favorable when one verification step prevents several later reasoning or invocation rounds; it is less pronounced when the task needs only a short tool chain.

Fig.~\ref{fig:llama_redundancy} summarizes the Llama redundancy analysis for the Llama backbone, where the goal is to estimate how much of the token budget is attributable to repeated or low-value context under the heuristic decomposition. We can note that up to 46.3\% unnecessary tokens can be reduced in GAIA.

\paragraph{\textbf{Ablation Study}}

\begin{figure}[t]
  \centering
  \includegraphics[width=0.7\linewidth]{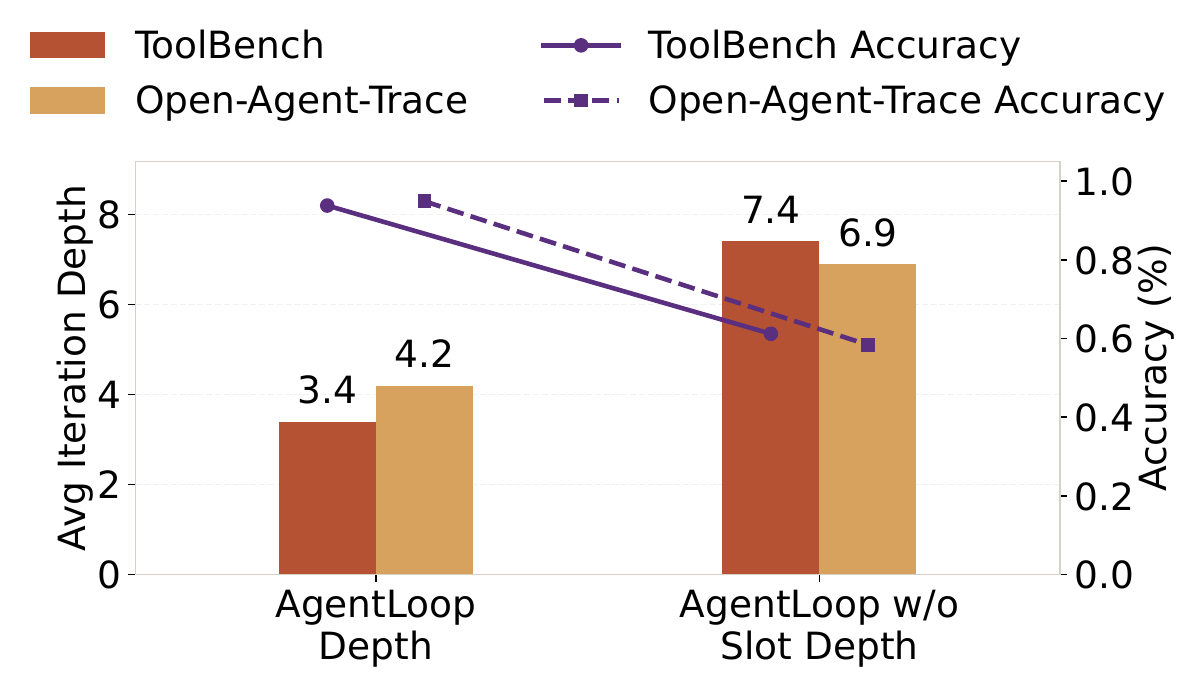}
  \caption{Ablation of the slot-centered control path by comparing full AgentLoop with AgentLoop w/o Slot.}
  \label{fig:ablation}
\end{figure}

To evaluate the slot-centered control path, Fig.~\ref{fig:ablation} compares the full system with AgentLoop w/o Slot and shows whether slot control improves efficiency without damaging answer quality. The ablation indicates fewer unproductive loops and higher final answer quality with slot-centered control.

The ablation keeps the same base model, tool environment, and dataset setting, but disables the slot-centered control path as a group, including the reflection score gate, convergence-round test, and answer integration/recheck step. It therefore measures the contribution of the coordinated slot-control path rather than the isolated effect of one switch. Without this path, average depth increases from about 3.4 to 7.43 and accuracy decreases from about 0.938 to 0.612 on ToolBench; on Open-Agent-Trace, depth increases from about 4.23 to 6.86 and accuracy decreases from about 0.949 to 0.584. These results show that coupling slot completion with Iteration Controller decisions improves the efficiency-quality trade-off.
The direction is consistent across both datasets: removing slot-centered control allows additional rounds without providing the same answer-quality support. The ablation thus attributes the improvement to explicit request-state tracking and its connection to the controller, rather than to a generic reduction in the number of calls.

\section{Discussion}

AgentLoop is intended as a runtime-state control layer rather than a replacement for existing agent frameworks. Its main value is most visible when an agent must coordinate multiple reasoning-action rounds and decide whether further execution will add useful information. In such settings, slot closure provides a task-level completion signal, while stability and low-value heuristics provide execution-level control signals. These signals complement model self-judgment, tool-call success, and workflow boundaries.

The results also define a practical deployment boundary. For tool-intensive or workflow-style requests, the additional state checks can reduce repeated context growth and move execution toward answer synthesis once the required information is stable. This makes AgentLoop suitable as an embedded control component inside broader service-computing agent systems.

Several limitations follow from how the control layer is currently built. The stability test that decides whether a state has converged is computed from lexical and set-level overlap, so a semantically equivalent paraphrase is still counted as a change and can permit one further round. The ablation in Fig.~\ref{fig:ablation} disables the slot-centered path as a group, which means the reported gap measures the combined effect of slot verification, the reflection score gate, and answer integration rather than the marginal value of any single component.

\section{Conclusion}

This paper presented AgentLoop, a runtime slot-closed execution control mechanism for tool-augmented LLM agents. The central view is that open-ended agent iteration can be converted into explicit runtime-state control: slot closure provides a task-level completion signal, while context stability, evidence stability, budget pressure, and low-value heuristics guide execution-level decisions. By coordinating Runtime State Manager, Slot State Verifier, and Iteration Controller, AgentLoop represents user requests as slot states, verifies evidence, attributes gaps, suppresses redundant context and evidence states, guards budget, and selects Continue Invocation, Answer Synthesis, or Terminate Iteration.

Experiments on thousands of requests from three datasets show that AgentLoop improves the balance between execution efficiency and answer quality when runtime context and tool evidence have room to stabilize. Across the two model backbones, AgentLoop reduces redundant context growth and high-depth execution, with total token cost reduced by up to 88.44\% and average depth reduced by up to 76.85\% against high-cost baselines. In practice, this saving is largely driven by fewer loop expansions and less trace carryover. The ablation further shows the role of slot-centered control: disabling it increases average depth from 3.4 to 7.43 on ToolBench and from 4.23 to 6.86 on Open-Agent-Trace, while reducing accuracy from 0.938 to 0.612 and from 0.949 to 0.584. Future work will explore lighter slot verification, finer-grained budget control, domain-specific slot extraction, and stability evaluation under concurrent service workloads.

\ifCLASSOPTIONcaptionsoff
  \newpage
\fi

\begingroup
\fontsize{7.6}{8.45}\selectfont
\setlength{\itemsep}{0pt}
\setlength{\parskip}{0pt}
\def\IEEEbibitemsep{0pt}
\let\oldthebibliography\thebibliography
\let\oldendthebibliography\endthebibliography
\renewenvironment{thebibliography}[1]{%
  \oldthebibliography{#1}%
  \fontsize{7.6}{8.45}\selectfont
  \setlength{\itemsep}{0pt}%
}{\oldendthebibliography}
\bibliographystyle{IEEEtran}
\IfFileExists{DeLoop/Bibliography.bib}{%
  \bibliography{IEEEabrv,DeLoop/Bibliography}%
}{%
  \bibliography{IEEEabrv,Bibliography}%
}
\endgroup

\begin{IEEEbiography}[{\includegraphics[width=1in,height=1.25in,clip,keepaspectratio]{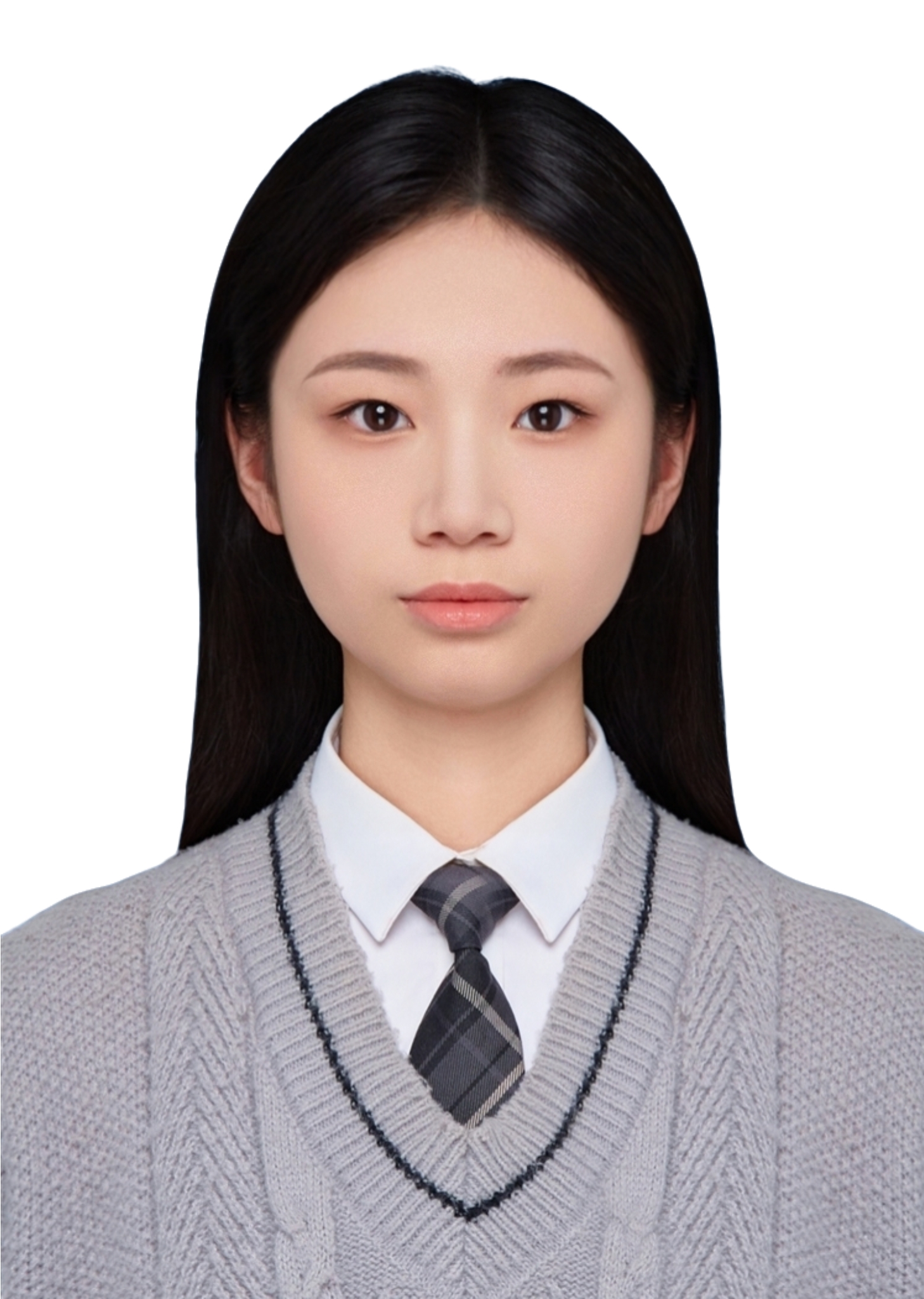}}]{Wanyi Zheng}
Wanyi Zheng received her Bachelor's degree from Hebei University in 2025. She is currently pursuing a Master's degree at the Southern University of Science and Technology and is also a joint-training student at the Shenzhen Institutes of Advanced Technology, Chinese Academy of Sciences. Her research interests include LLM-based agent systems and intelligent service systems, with a primary focus on agent runtime optimization, efficient tool use, and system-level optimization for LLM-powered agents.
\end{IEEEbiography}
\vspace{-4.0em}

\begin{IEEEbiography}[{\includegraphics[width=1in,height=1.25in,clip,keepaspectratio]{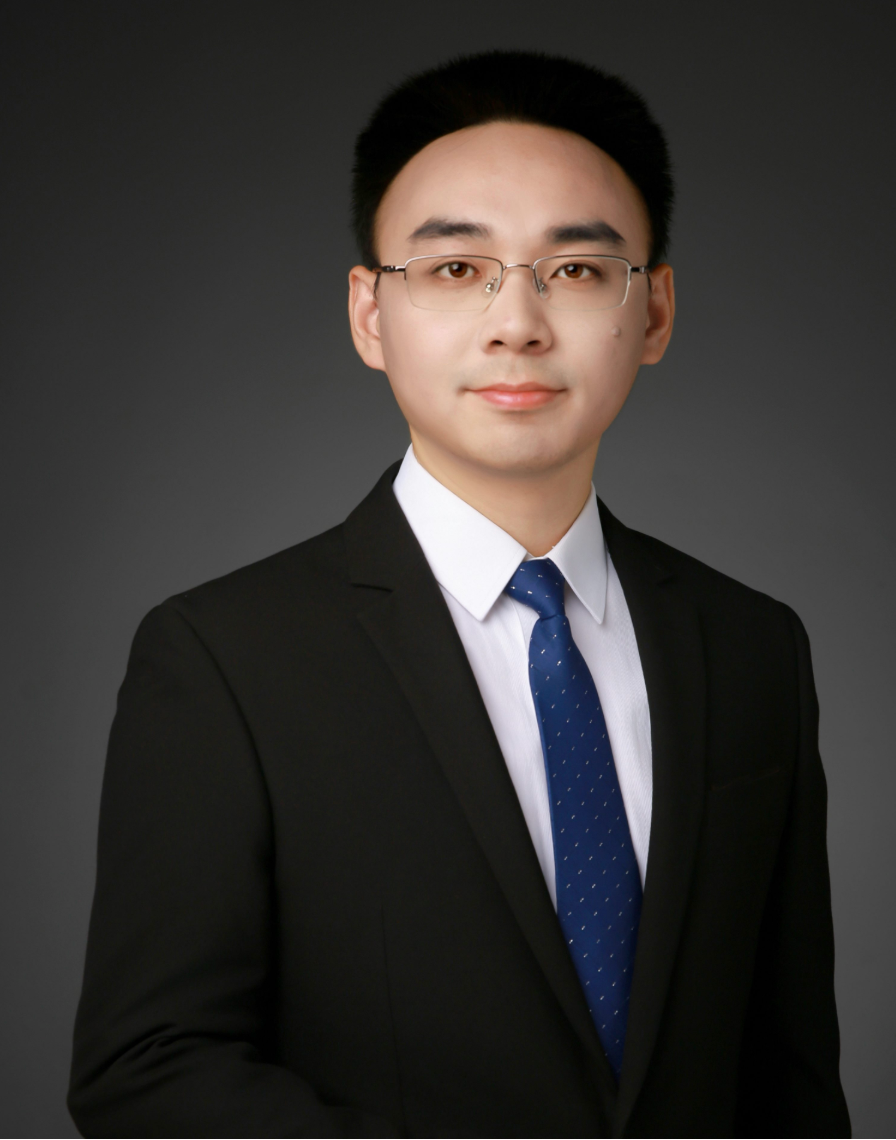}}]{Minxian Xu}
 (Senior Member, IEEE) received the Ph.D. degree from the University of
Melbourne, Melbourne, VIC, Australia, in 2019.
He is currently an Associate Professor with Shenzhen Institutes of Advanced Technology, Chinese
Academy of Sciences. His research interests include
cloud-native systems for LLM inference and AI
infrastructure. He has co-authored more than 90
peer-reviewed papers in prominent journals and
conferences, including ACM CSUR,
IEEE TSC, TC, TMC, TAAS, TCC, TOIT, ICSOC and ICWS, thes work attracted 7,000+ citations. He serves as the Associate Editor of IEEE TSC. 
\end{IEEEbiography}
\vspace{-4.0em}

\begin{IEEEbiography}[{\includegraphics[width=1in,height=1.25in,clip,keepaspectratio]{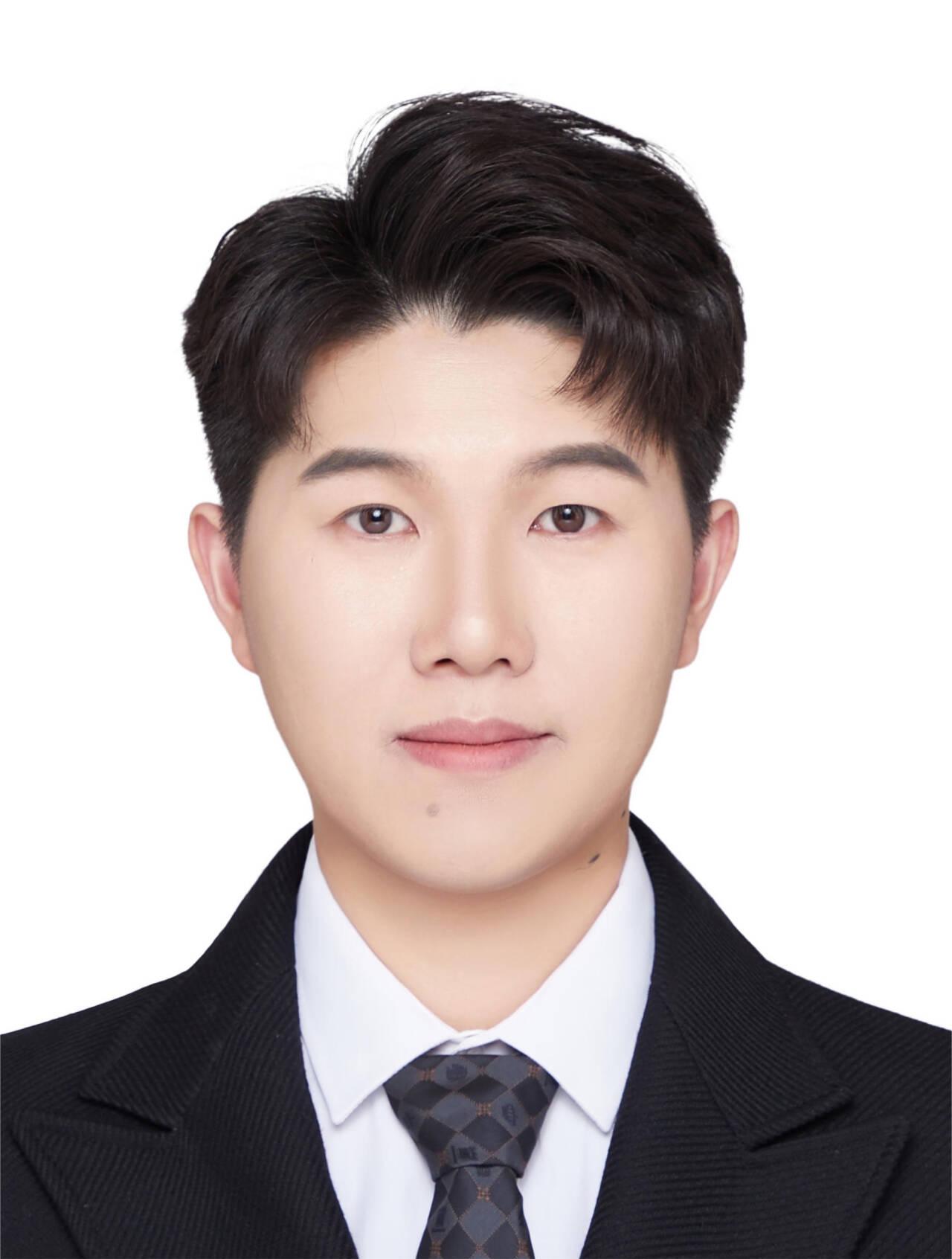}}]{Kan Hu}
 received his Master's degree from the University of Chinese Academy of Sciences in 2025. He has been awarded a PhD scholarship at the University of Melbourne and is yet to commence his doctoral studies. His research interests include LLM-based agent systems, and he particularly concentrates on optimizing large-scale multi-agent systems by leveraging cloud computing paradigms—especially in terms of resource elasticity, load balancing, and efficient task coordination, aiming to enhance system performance.
\end{IEEEbiography}
\vspace{-4.0em}

\begin{IEEEbiography}[{\includegraphics[width=1in,height=1.25in,clip,keepaspectratio]{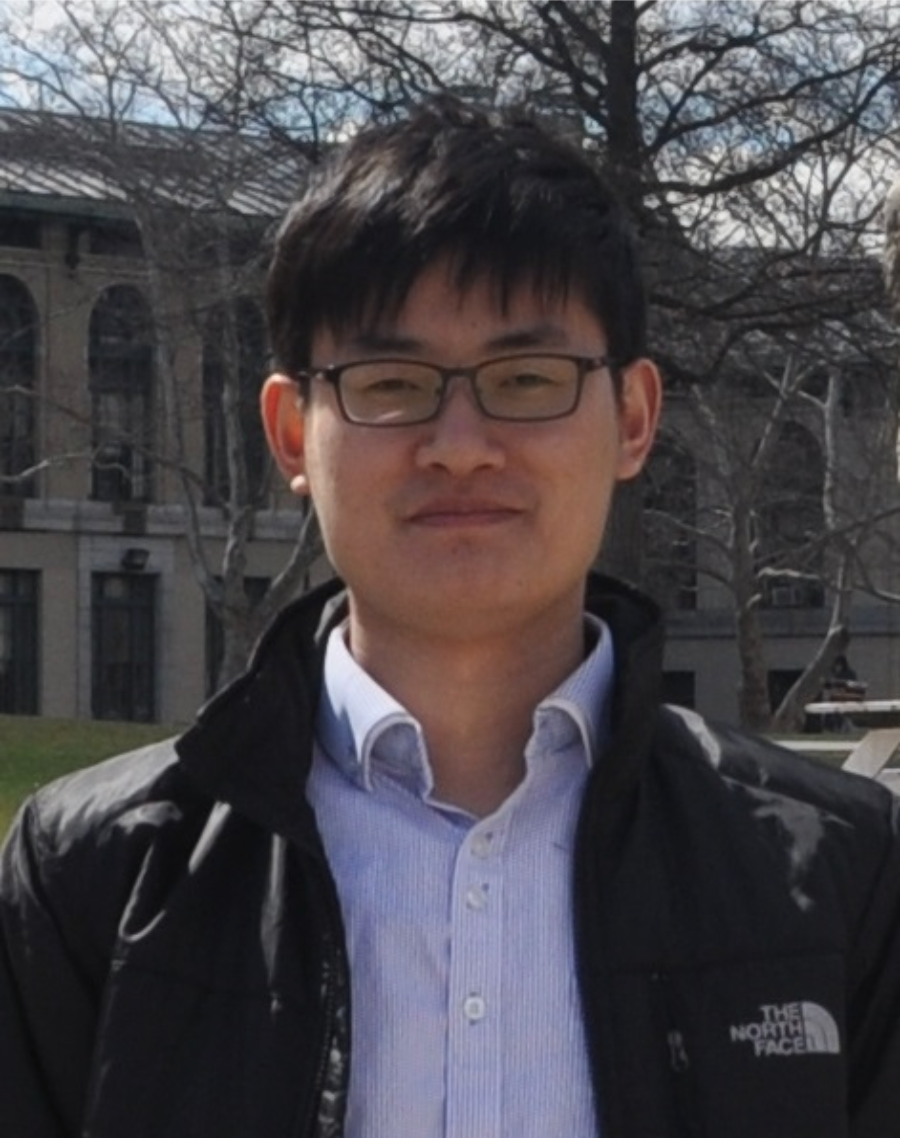}}]{Kejiang Ye}
 (Senior Member, IEEE) received the B.S. and Ph.D. degrees from Zhejiang University. He is currently a Professor and the Director of the Research Center for Cloud Computing, Shenzhen Institutes of Advanced Technology, Chinese Academy of Sciences. He was a Postdoctoral Research Associate with Carnegie Mellon University, Pittsburgh, PA, USA. His research interests include digital technology and systems, such as cloud computing, big data, and industrial Internet. He is a Distinguished Member of the China Computer Federation.
\end{IEEEbiography}
\vspace{-4.0em}

\begin{IEEEbiography}[{\includegraphics[width=1in,height=1.25in,clip,keepaspectratio]{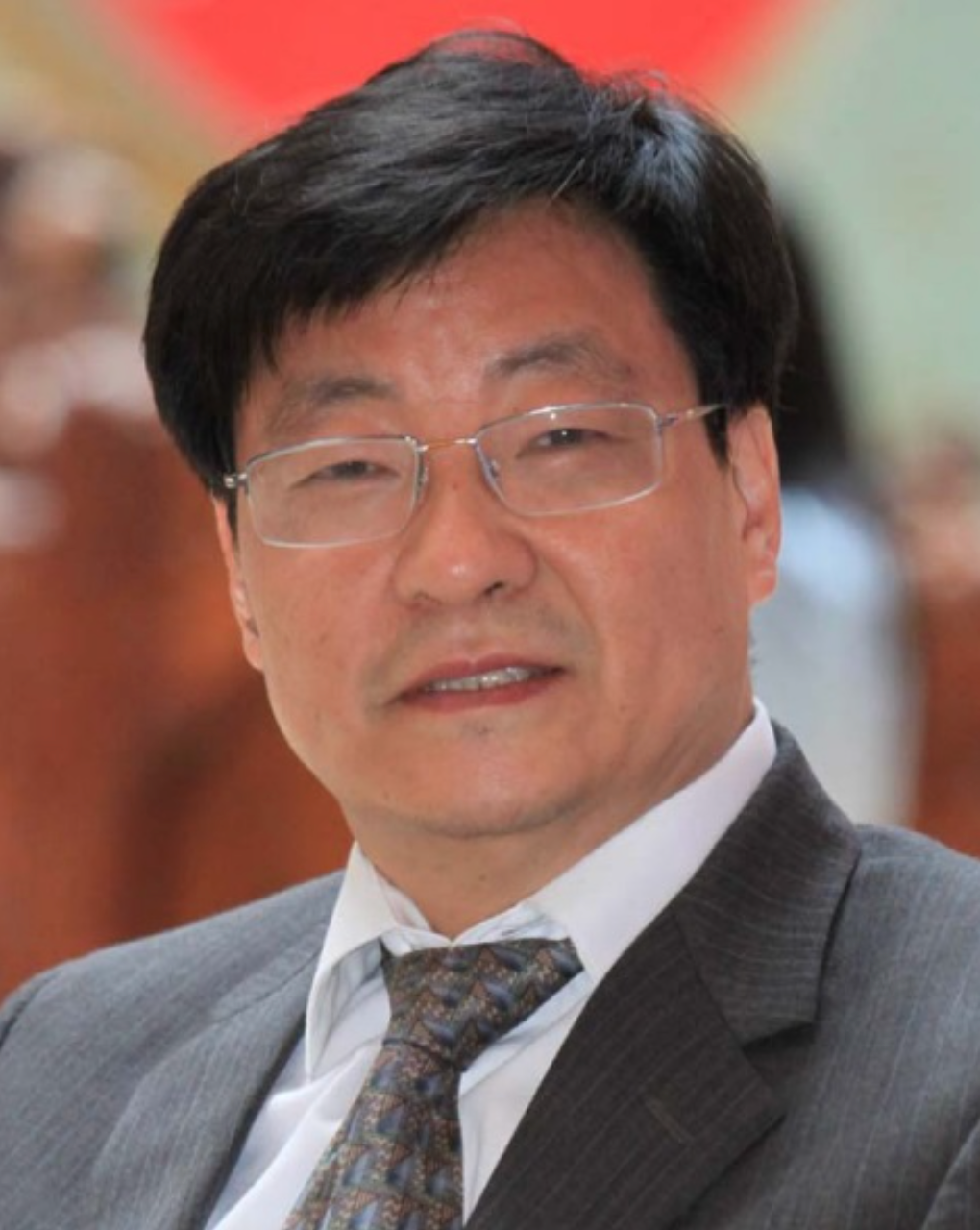}}]{Chengzhong Xu}
 (Fellow, IEEE) received the Ph.D. degree in computer science and engineering from The University of Hong Kong, Hong Kong, in 1993. He is currently the Dean of the Faculty of Science and Technology and the Interim Director of the Institute of Collaborative Innovation, University of Macau. His research interests include parallel and distributed computing, with an emphasis on resource management for performance, reliability, availability, power efficiency, and security. His work spans servers and cloud datacenters, wireless embedded devices, and edge AI systems, with applications in smart city and autonomous driving. He has authored two research monographs and more than 600 papers, which have received more than 22,000 citations with an H-index of 77, and have been cited in more than 300 international patents.
\end{IEEEbiography}
\end{document}